\documentclass[lettersize,journal]{IEEEtran}
\usepackage{amsmath,amsfonts}
\usepackage{algorithmic}
\usepackage{algorithm}
\usepackage{array}
\usepackage[caption=false,font=normalsize,labelfont=sf,textfont=sf]{subfig}
\usepackage{textcomp}
\usepackage{stfloats}
\usepackage{url}
\usepackage{verbatim}
\usepackage{graphicx}
\usepackage{cite}
\usepackage{float}
\newfloat{algorithm}{t}{lop}
\usepackage{epsfig}
\usepackage{bbm}
\usepackage{makecell}
\DeclareMathOperator*{\argmin}{argmin}
\DeclareMathOperator*{\argmax}{argmax}
\usepackage{comment}

\begin{document}

\title{Can Perceptual Guidance Lead to Semantically Explainable Adversarial Perturbations?}

\author{Charantej Reddy Pochimireddy, Aditya T. Siripuram, Sumohana S. Channappayya~\IEEEmembership{Senior Member,~IEEE}
}

\markboth{Journal of \LaTeX\ Class Files,~Vol.~14, No.~8, August~2021}%
{Shell \MakeLowercase{\textit{et al.}}: A Sample Article Using IEEEtran.cls for IEEE Journals}


\maketitle

\begin{abstract}
It is well known that carefully crafted imperceptible perturbations can cause state-of-the-art deep learning classification models to misclassify. Understanding and analyzing these adversarial perturbations play a crucial role in the design of robust convolutional neural networks. However, their mechanics are not well understood. In this work, we attempt to understand the mechanics by systematically answering the following question: do imperceptible adversarial perturbations focus on changing the regions of the image that are important for classification? In other words, are imperceptible adversarial perturbations semantically explainable? Most current methods use $l_p$ distance to generate and characterize the imperceptibility of the adversarial perturbations. However, since $l_p$ distances only measure the pixel to pixel distances and do not consider the structure in the image, these methods do not provide a satisfactory answer to the above question. To address this issue, we propose a novel framework for generating adversarial perturbations by explicitly incorporating a ``perceptual quality ball" constraint in our formulation. Specifically, we pose the adversarial example generation problem as a tractable convex optimization problem, with constraints taken from a mathematically amenable variant of the popular SSIM index. We use the MobileNetV2 network trained on the ImageNet dataset for our experiments. By comparing the SSIM maps generated by our method with class activation maps, we show that the perceptually guided perturbations introduce changes specifically in the regions that contribute to classification decisions i.e., these perturbations are semantically explainable. 
\end{abstract}

\begin{IEEEkeywords}
SSIM, Adversarial perturbations, Explainability. 
\end{IEEEkeywords}
\section{Introduction}
\IEEEPARstart{T}{he} fields of Artificial Intelligence (AI) and Machine Learning (ML) have seen tremendous growth and development in the last decade that is expected to have a major impact on humankind in the foreseeable future. This growth can largely be attributed to advances in deep learning, which in turn can be attributed to the availability of large data sets and efficient hardware. These advances allow us to accurately train highly complex deep learning models which are used in a myriad of application areas, including computer vision, speech and audio processing, natural language processing, medical imaging, and finance, to name a few.

In theory, neural networks can represent/approximate any real-valued function (or at least a wide variety of functions) when appropriate weights and architecture are chosen according to the \textit{Universal Approximation Theorem} \cite{cybenko1989approximation}. With sufficient data and the right choice of architecture, one can achieve state-of-the-art results on a variety of machine learning problems using deep neural networks. 

Analyzing the robustness of deep learning models to adversarial inputs has received significant attention over the past few years. A majority of these works can be classified into four broad areas: 1) generating adversarial examples/crafting adversarial perturbations, 2) defending against such `attacks' \cite{yuan2019adversarial,qiu2019review}, 3) evaluation and certification \cite{wong2017provable, raghunathan2018semidefinite}, and 4) interpretability of adversarial examples/perturbations \cite{ilyas2019adversarial}. However, we still lack a clear understanding of the key underlying factors for the existence of this phenomenon. Because of this, it is challenging to design defence mechanisms that work for all kinds of adversarial attacks/perturbations. The adversarial example generation problem is highly non-convex, and heuristic or other approximation techniques are required to solve it. 

In their seminal work, Szegedy et al. introduced adversarial examples in \cite{szegedy2013intriguing} that are crafted by solving an optimization. 
Later, faster adversarial perturbation methods with closed-form solution (direct expression to generate perturbations) were proposed by using the norm bounded constraints; examples include FGSM ($l_\infty$ bounded perturbations) \cite{goodfellow2014explaining}, FGM ($l_2$ bounded perturbations) \cite{dong2018boosting}, and PGD ($l_\infty$ and $l_2$ bounded perturbations). Several attack and defense algorithms have since been proposed, and many of these methods are summarized in  \cite{yuan2019adversarial,qiu2019review}. Most existing defences are not completely robust: adversarial attacks/perturbations can be specifically engineered to target the said defense technique. This has led to a self-sustaining cycle - with attacks leading to defenses, and defenses paving the way for new attacks. Despite the many results in this area, a complete understanding of the adversarial perturbation landscape is still lacking. Some proposed hypotheses for explaining adversarial perturbations include high dimensional spaces \cite{goodfellow2014explaining}, data in-completion, model capacity \cite{yuan2019adversarial}, and the presence of highly predictive but non-robust features (spurious correlations) \cite{ilyas2019adversarial}. The authors in \cite{hendrycks2018benchmarking} proposed that the distribution shift between training and test data set, combined with the high dimensional continuous data space, as the key reasons for adversarial examples. 

Along these lines, our interest is not to create another attack but rather to understand the adversarial landscape, primarily with respect to perceptual similarity. Our investigation aims to systematically explore the semantic significance of the regions affected by adversarial perturbations. In particular, we try to answer the following question: \emph{do imperceptible adversarial perturbations focus on changing the regions of the image that are important for classification? In other words, are imperceptible adversarial perturbations semantically explainable? }

In most of the adversarial example attacks, $l_p$ distance metrics are used for crafting the adversarial perturbations. However, it is well-known that $l_p$ distance metrics are not good at measuring perceptual similarity between two images \cite{wang2004image}, and that these metrics are neither necessary nor sufficient for perceptual similarity \cite{sharif2018suitability}. For example, a mean shift in the image results in a high $l_p$ distance, but perceptually, the mean shifted image remains close to the original image. For the perturbations to be imperceptible, the original and modified image must be similar in some metric that respects structural (perceptual) similarity. Thus to answer the question above, we use the Structural Similarity (SSIM) index \cite{wang2004image} to generate the adversarial perturbations instead of $l_p$ distance metrics. The SSIM index is a popular measure to evaluate the perceptual similarity/quality of images. It is a full-reference image quality metric that measures the quality of a distorted image with respect to the ground-truth pristine image. Compared to traditional $l_p$ distance metrics, the SSIM index is known to correlate better with the human perception of image quality/distortion. The following are the main contributions of this work:


\begin{enumerate}
 \item In order to address the question posed earlier, we propose a perceptually guided adversarial example generation technique by leveraging the useful mathematical properties of the SSIM index. The SSIM metric is analyzed in depth in order to construct a novel convex formulation (which we call perceptually guided adversarial perturbation (PGAP)) to generate the adversarial perturbations. We also provide a closed-form approximation (called Faster PGAP (FPGAP)) to the proposed convex problem.
 \item We then try to systematically investigate (both qualitatively and quantitatively) adversarial examples and their relationship to semantically significant regions of the image.  We do this by comparing the perturbations generated using the SSIM index with class activation maps generated using GradCAM++ \cite{chattopadhay2018grad}. We quantify this comparison using IOU-based metrics and precision; and observe that compared to other norm-bounded approaches, our method gives about a factor of 2-3 higher precision scores (Fig \ref{fig:iou_metrics}).
     \item We thus conclusively answer the question posed in the introduction. Adversarial perturbations generated by incorporating an SSIM ball constraint (instead of the $l_p$ ball constraint as in other works) seem to be changing only the regions of the image significant for classification, at least to a degree much higher than other norm-bounded attack methods.
     \item In order to validate our adversarial example generation method, we also establish that our method gives a much higher fooling rate at a given average SSIM compared to other similar methods.
\end{enumerate}
We find it very intriguing that a \emph{perceptually-aware} formulation makes the adversarial example generation \emph{semantically-aware}.
\section{Related work}
The SSIM index has been considered previously in the adversarial perturbation setting, and we briefly summarize two relevant methods next. In \cite{jordan2019quantifying}, the authors proposed stronger attacks by combining different types of attacks (adding adversarial noise, rotating, translating, or performing spatial transformations on images) and used the SSIM index to quantify the strength of the adversarial attack. In \cite{rozsa2016adversarial}, the authors introduced a new measure called Perceptual Adversarial Similarity Score (PASS) using the SSIM index to quantify the adversarial examples and use the PASS in the process of generating adversarial examples. 

Our contribution differs significantly from both of these methods, which we outline below. By imposing constraints from a mathematically amenable variant of the SSIM index \cite{brunet2011mathematical} (and exploiting useful properties like quasi convexity), and by taking a suitable approximation of the loss function, we pose the adversarial example generation problem as a quadratically constrained quadratic program (QCQP). We also provide a closed-form solution to the approximation of the QCQP that allows for faster implementation. In fact, none of the existing adversarial
example techniques that use image quality metrics provide a closed-form solution to \eqref{eq:formulation}. We not only give a convex formulation \eqref{eq:opt_ssim_s2}, but also provide a closed-form approximation \eqref{eq:cf_min}. Thus our method is in line with other norm bounded techniques (PGD, FGSM, etc.), which employ a closed-form solution to generate adversarial perturbations. The perturbations we generate are not necessarily additive, nor are they obtained by a parameterized rotation or spatial transformation; we propose a model-free technique to generate adversarial examples that are structurally and thereby perceptually similar to the original image.

While this manuscript was in development, we also became aware of another line of work \cite{hameed2021perceptually} that uses SSIM to generate adversarial examples. However, our work differs in the following key aspects: First, our goal, as opposed to the work in \cite{hameed2021perceptually}, is to investigate the question posed in the introduction, and not simply to generate another attack. As such, we have a detailed and systematic analysis of the regions of perturbations as compared to the regions identified by GradCAM++, which are not present in \cite{hameed2021perceptually}. Secondly, we provide a tractable convex formulation of the adversarial example generation problem. We use the analysis from \cite{brunet2011mathematical} primarily to achieve this goal. Thirdly, in addition to the convex formulation, we also provide a fast approximation to the convex formulation (FPGAP), which brings our technique (computationally and structurally) in line with existing norm-bounded methods. This is in contrast to the gradient-descent-based Lagrangian optimization used in \cite{hameed2021perceptually} (as a consequence, our method also does not need any additional hyper parameters). This fast approximation also allows us to perform experiments on datasets like ImageNet, which would not be possible with the initial formulation. Indeed such experiments are not done in \cite{hameed2021perceptually}. Finally, we also provide a systematic analysis of the comparative impact of the hyper-parameters involved in our algorithm ($\epsilon_1$ and $\epsilon_2$ in \eqref{eq:opt_ssim_s1_s2}).


\subsection{Organization}
The paper is organized as follows: in section \ref{sec: problem setup}
we briefly discuss the standard norm bounded adversarial example generation methods and formulate an optimization problem \eqref{eq:opt_ssim}. In section \ref{sec:ssim_index}, we explain the SSIM index along with its mathematically amenable variant and discuss some of its properties which will be used to generate adversarial examples using SSIM constraints. In section \ref{sec:aeGeneration_ssim}, we use ideas from \ref{sec:ssim_index} to modify \eqref{eq:opt_ssim} to formulate an optimization problem \eqref{eq:opt_ssim_s2} to generate adversarial examples using SSIM index. We also propose an approximate solution \eqref{eq:cf_min} to this optimization problem which enables the faster generation of adversarial examples. In section \ref{sec:results}, we analyze the norm bounded methods along with the proposed method qualitatively and quantitatively and present some additional results on the perceptual quality of adversarial examples generated by different methods. We conclude the paper with some closing remarks in section \ref{sec:conclusion}.

\section{Problem Setup}
\label{sec: problem setup}
Suppose $\mathcal{D}$ represents the data set with entries $(x_i,y_i)$, where $x_i$ represents the $i^{th}$ data point and $y_i$ represents the corresponding label. Let $f$ denote the machine learning model and its prediction $\hat{y}$ (i.e., $\hat{y}_i = f(x_i)$). In the supervised learning framework we try to minimize the loss function $\mathcal{L}$ with respect to given data set $\mathcal{D}$ and find the best suited model parameters $w$.
\begin{equation*}
\begin{aligned}
\min_{w} \quad & \sum_{i}^{}{\mathcal{L}(w,x_i,y_i)}.
\end{aligned}
\end{equation*}
An adversarial example $x_{adv}$ is similar (in some metric) to a data point $x$ in the data set $\mathcal{D}$, such that the machine learning models misclassifies the input (i.e., $f(x_{adv}) \neq f(x)$). Finding an adversarial example at a given data point $x$ with label $y$ can be formulated as an optimization problem based on loss function:
In this formulation, we find $x_{adv}$ close to a given data point $x$, which maximizes the loss function 
\begin{equation}
\label{eq:formulation}
\argmax_{x_{adv}} \ \mathcal{L}(w,x_{adv},y) \quad
\textrm{s.t.} \quad  d(x,x_{adv}) \leq \epsilon.
\end{equation}

Here $\epsilon$ is the allowed level of perturbation. We refer to $x_{adv}$ as adversarial example and $x-x_{adv}$ as adversarial perturbation.

Note that the feasibility region of the optimization problem above \eqref{eq:formulation} varies based on the value of $\epsilon$. The formulation in \eqref{eq:formulation} is similar to the inner maximization problem from \cite{madry2017towards}. The distance metric $d(x,x_{adv})$ is typically an $l_p$ distance \[d(x,x_{adv}) = ||x-x_{adv}||_p \text{ where } ||a||_p = \left(\sum_{i}^{}|a_i|^p\right)^{\frac{1}{p}}. \] 

Typically, a linear approximation of the loss function around the data point $x$ is used: 
\begin{equation*}
 \mathcal{L}(w,x_{adv},y) \approx \mathcal{L}(w,x,y) + (x_{adv}-x)^{T} \nabla_{x} \mathcal{L}(w,x,y).
\end{equation*}
So to find $x_{adv}$ we maximize the second term above. Note that since 
\[
\argmax_{x_{adv}}\  \mathcal{L}(w,x_{adv},y) \approx \argmax_{x_{adv}}  (x_{adv})^{T} \nabla_{x} \mathcal{L}(w,x,y),
\]
we can just maximize the inner product between $x_{adv}$ and $\nabla_{x} \mathcal{L}(w,x,y)$.  Also note that the gradients $\nabla_{x} \mathcal{L}(w,x,y)$ can be readily extracted from the model.

Important distance metrics used in the literature are $l_\infty$, which measures the maximum absolute change in the pixel values \cite{goodfellow2014explaining}, \cite{kurakin2016adversarial}, \cite{carlini2017towards}; $l_2$ which measures the Euclidean distance of change in the pixel values~\cite{carlini2017towards},~\cite{dong2018boosting}; $l_1$ measures the total absolute change in the pixel values~\cite{chen2018ead}, and $l_0$ measures the number of pixels that differ~\cite{carlini2017towards},~\cite{su2019one},~\cite{xu2018structured}. 

As discussed earlier, $l_p$ distance metrics are not good at measuring perceptual similarity between two images. If we consider
\[
d(x,x_{adv})=\sqrt{1-\text{SSIM}(x,x_{adv})},
\] then the optimization problem for generating adversarial example becomes:  
\begin{equation}
\label{eq:opt_ssim}
\begin{aligned}
\argmax_{x_{adv}} \quad & (x_{adv})^{T} \nabla_{x} \mathcal{L}(w,x,y) \\
\textrm{s.t.} \quad & \text{SSIM}(x,x_{adv}) \geq 1-\epsilon^2
\end{aligned}
\end{equation}

However, the metric $d(x,x_{adv})=\sqrt{1-\text{SSIM}}$ defined above is non-convex, and so the above problem can become intractable. So we use a variant of the SSIM index \cite{brunet2011mathematical} to generate imperceptible adversarial perturbations. We review these ideas next. Readers familiar with the SSIM index can skip to Section \ref{sec:aeGeneration_ssim}.

\subsection{Structural Similarity (SSIM) Index}
\label{sec:ssim_index}
The SSIM index between the two image patches $X$ and $Y$ is computed using a combination of three distortion measurement components: luminance ($l$), contrast ($c$), and structure/correlation ($s$), that are defined as follows.
\[
l = \frac{2\mu_X\mu_Y+c_1}{\mu_x^2+\mu_Y^2+c_1},\quad c = \frac{s_{X}s_{Y}+c_2}{s_{X}^2+s_{Y}^2+c_2}\quad  s = \frac{s_{X,Y}+c_3}{s_{X}s_{Y}+c_3},
\]
 where $\mu_{X}, \mu_{Y}$ represent the mean of $X$ and $Y$ respectively, $s_{X}^2, s_{Y}^2$ represent the variances of $X$ and $Y$ respectively, and $s_{X,Y}$ represents the co-variance between the $X$ and $Y$. Here, $c_1, c_2$ and $c_3$ are small numerical constants that ensure numerical stability when the denominators are close to zero. We can also say that these constants aim to characterize the saturation effects of the visual system at low luminance and contrast regions. The first two terms $l$ and $c$ measure nonstructural distortion, while the last term $s$ measures structural distortion (or absence of correlation) between the two images. The structural similarity or SSIM between the images $X$ and $Y$ is defined as the product of the luminance, contrast and structure terms defined above, i.e., $\text{SSIM}(X,Y) =  l.c.s$.


An SSIM quality map is constructed by computing the SSIM index between pairs of corresponding local patches in the two images, and the overall SSIM index is computed by averaging the patch level values in the SSIM map.

While the SSIM index is indeed a better method compared to MSE for measuring perceptual similarity between two images, it does not satisfy the triangle inequality and thus is not a distance metric, limiting its use in convex problem formulations. However, the SSIM index can be converted to a normalized root mean square error (NRMSE) measure, which is a valid distance metric \cite{brunet2011mathematical}. The square of such a metric is not convex but is locally convex, and quasi-convex \cite{brunet2011mathematical}, thereby making the SSIM index a feasible target for optimization. We use these insights for our problem formulation. Next, we briefly review these ideas, develop the notation.

In the standard form of the SSIM index, we set the numerical constants $c_3 = c_2/2$, resulting in the SSIM having only two terms 
\begin{equation} 
    \text{SSIM}(X,Y) = S_1(X,Y) \hspace{0.2cm} S_2(X,Y),
    \label{eq:math_ssim}
\end{equation} 
where $S_1=l$ and 
 \begin{equation*}
     S_2 = c.s = \left(2s_{X,Y}+c_2\right)/\left(s_{X}^2+s_{Y}^2+c_2\right).
 \end{equation*}
 
It can be seen that $\sqrt{1-\text{SSIM}}$ is not a metric but $\sqrt{1-S_1}$ and $\sqrt{1-S_2}$ are normalized metrics \cite{brunet2011mathematical}. Now we set $d_1 = \sqrt{1-S_1}$ and $d_2 = \sqrt{1-S_2}$, and the vector $d = [d_1, d_2]$.  It can be seen that $d$ is a vector of normalized metrics obtained from the root mean square error \cite{brunet2011mathematical}.

 The SSIM index can be approximated with the vector of metrics $d(X,Y)$ as 
 \begin{equation}
 \begin{split}
     ||d(X,Y)||_2 & = \sqrt{(d_1)^2+ (d_2)^2}  = \sqrt{2-S_1-S_2}.
 \end{split}
 \end{equation}
 
 We note that
 \begin{equation}
     \sqrt{1-\text{SSIM}}  = \sqrt{1-S_1S_2} = \sqrt{d_1^2+d_2^2-d_1^2d_2^2}.
 \end{equation}
 
 We observe that $||d(X,Y)||_2$ serves as a lower order approximation of $\sqrt{1-\text{SSIM}}$. We can also write 
 \begin{equation} 
 \begin{split}
     S_1 & = 1-\text{NMSE}(\mu_{X},\mu_{Y},c_1) \\
     S_2 & = 1-\text{NMSE}(X-\mu_{X},Y-\mu_{Y},c_2),
 \end{split}
 \label{eq:s_1_2}
 \end{equation}
 where $\text{NMSE}$ is the normalized mean squared error given by
 \begin{equation}
     \text{NMSE}(X, Y, c) = \frac{||X-Y||^2}{||X||^2+||Y||^2+c}.
     \label{eq:nmse}
 \end{equation}
 We use these ideas to modify problem \eqref{eq:opt_ssim}.
 
 \section{Adversarial example generation using SSIM}
 \label{sec:aeGeneration_ssim}
 
 \begin{figure*}[!t]
\centering
\subfloat[]{\includegraphics[width=3in]{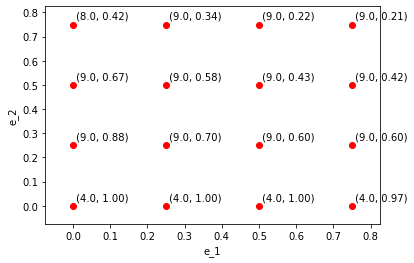}}%
\hfil
\subfloat[]{\includegraphics[width=3in]{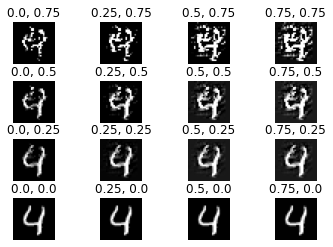}}%
\caption{Image on the left: x-axis and y- axis correspond to $\epsilon_1$ and $\epsilon_2$ respectively, at each highlighted point on the grid has label with first co-ordinate as model prediction and second co-ordinate as ssim value. Image on the right: corresponding images with $\epsilon_1$ and $\epsilon_2$ mentioned on top of them.}
\label{fig:grid_e1_e2}
\end{figure*}

The structural similarly index can be written as a product of two terms i.e., $\text{SSIM}=S_1.S_2$ where $S_1$ captures the luminance similarity, and $S_2$ captures the structure and contrast similarity. From \cite{brunet2011mathematical} (summarized in Section \ref{sec:ssim_index} above) we know that $S_1$ and $S_2$ have appealing convexity properties. Going back to the problem formulation of \eqref{eq:opt_ssim}, we see that the condition on $\text{SSIM}(x,x_{adv})$ in equation \eqref{eq:opt_ssim} can be replaced by conditions on $S_1(x,x_{adv})$ and $S_2(x,x_{adv})$, as discussed in \eqref{eq:math_ssim}. By choosing $\epsilon_1$ and $\epsilon_2$ suitably, we may rewrite the optimization problem from \eqref{eq:opt_ssim} as
\begin{equation}
\begin{aligned}
\argmax_{x_{adv}} \quad & (x_{adv})^{T} \nabla_{x} \mathcal{L}(w,x,y) \\
\textrm{s.t.} \quad & S_1(x,x_{adv}) \geq 1-\epsilon_1^2 \\
& S_2(x,x_{adv}) \geq 1-\epsilon_2^2
\end{aligned}
\label{eq:opt_ssim_s1_s2}
\end{equation}

We analyse the constraints of \eqref{eq:opt_ssim_s1_s2} in more detail in Appendix \ref{sec:constraints_appendix}. The first constraint $S_1(x,x_{adv}) \geq 1-\epsilon_1^2 $ is a linear constraint (this forces $x_{adv}$ to lie in an intersection of two half-spaces). The second constraint in \eqref{eq:opt_ssim_s1_s2} $S_2(x,x_{adv}) \geq 1-\epsilon_2^2$ is a quadratic constraint (this forces $x_{adv}$ to be in a high dimensional sphere). Based on the constraints \eqref{eq:c1}, \eqref{eq:c2}, and the objective, the optimization problem in \eqref{eq:opt_ssim_s1_s2} is convex; in particular it is a Quadratically Constrained Quadratic Program (QCQP).

Conceptually, Constraint 1 corresponds to non-structural perceptual features (luminance), and Constraint 2 corresponds to structural/perceptual features in the image. The parameters $\epsilon_1$ and $\epsilon_2$ fix the allowed tolerances in these features and decide the feasibility region of the optimization problem. 

\begin{figure}
\centering
\subfloat[]{\includegraphics[width=1.2in]{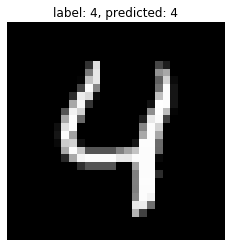}}%
\hfil
\subfloat[]{\includegraphics[width=1.2in]{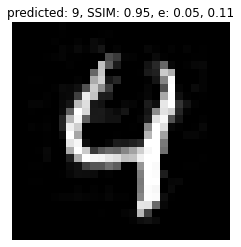}}%
\caption{Image on the left: original test image with label $y = 4$ and model prediction $\hat{y} = 4$. Image on the right: adversarial example generated by solving \eqref{eq:opt_ssim_s1_s2} ($\epsilon_1 = 0.05$ and $\epsilon_2 = 0.11$) with perceptual quality 0.95 (SSIM).}
\label{fig:ae_auto}
\end{figure}
To understand the relative impact of $\epsilon_1$ and $\epsilon_2$, we take an image with label $y = 4$ from the MNIST digits test data set and generate adversarial examples for different values of $\epsilon_1$ and $\epsilon_2$. For this exercise, we use a CNN model, which has around $99\%$ accuracy on the test data. In Figure \ref{fig:grid_e1_e2}, the image on the left shows the model prediction of optimization problem output and its SSIM index with respect to original image, as a function of $\epsilon_1$ and $\epsilon_2$. On the right, the corresponding output images are shown. From Figure \ref{fig:grid_e1_e2} one important observation that can be made is that the impact of $\epsilon_2$ (Constraint 2) is more on the solution (model prediction and perceptual quality of optimization output) compared to the $\epsilon_1$ (Constraint 1). This observation can also be validated by calculating the corresponding optimal dual variables. For the above example in Figure \ref{fig:ae_auto}, the dual variable corresponding to Constraint 2 is around 145 times larger compared to the dual variable of Constraint 1. Since $\epsilon_2$ has significantly higher impact compared to $\epsilon_1$, the algorithm can be simplified by taking $\epsilon_1 = 0$, effectively removing the Constraint 1 above. This modification results in solving the following optimization problem: 
\begin{equation}
\label{eq:opt_ssim_s2}
\begin{aligned}
\argmax_{x_{adv}} \quad & (x_{adv})^{T} \nabla_{x} \mathcal{L}(w,x,y) \\
\textrm{s.t.} \quad & \mu_{{x}_{adv}} = \mu_{x} \\
& S_2(x,x_{adv}) \geq 1-\epsilon_2^2
\end{aligned}
\end{equation}
\subsection{Proposed Method -- Perceptually Guided Adversarial Perturbation (PGAP)}
In practice, solving \eqref{eq:opt_ssim_s2} may not lead to an adversarial example. This is because the formulation of \eqref{eq:opt_ssim_s1_s2} assumed a linear approximation to the loss function, which may be accurate. Along the lines of \cite{madry2017towards}, we propose an iterative technique that repeatedly solves \eqref{eq:opt_ssim_s2}.
We first fix an $\epsilon_2$ and solve
\eqref{eq:opt_ssim_s2}. Note that at this point the obtained solution $x_{adv}$ may not be adversarial (i.e., we may not have $f(x_{adv}) \neq f(x)$). We recalculate the gradients at the obtained $x_{adv}$ and solve \eqref{eq:opt_ssim_s2} using the updated gradient. This process is repeated until an adversarial example is found (see Algorithm \ref{alg:alg_iter_AE} for a summary). The iterative approach helps in overcoming the limitations imposed by the linear approximation to the loss function.


One drawback of the proposed method (Algorithm \ref{alg:alg_iter_AE}) is that the QCQP is slow to solve on large datasets (for e.g., ImageNet). Hence next, we present next a faster algorithm that uses an efficient approximation of the solution to \eqref{eq:opt_ssim_s2}.

\begin{algorithm}[ht]
\caption{Adversarial example generation}
\label{alg:alg_iter_AE}
  \begin{algorithmic}
  \STATE
  \STATE {\textsc{PGAP}}($x, \text{model}, \text{label},\epsilon_2, \text{iterNum}$)
  \STATE \hspace{0.5cm} $i = 0$
  \STATE \hspace{0.5cm}  $\mathbf{While}$ {$i \leq \text{iterNum}$} $\mathbf{do}$
  \STATE \hspace{0.5cm} $\displaystyle x_{adv} \leftarrow \argmax_{x_{adv}} \hspace{0.1cm} (x_{adv})^{T} \nabla_{x} \mathcal{L}(w,x,y) \hspace{0.2cm} $
  \STATE \hspace{1.75cm} $ \textrm{s.t.} \hspace{0.2cm} \mu_{{x}_{adv}} = \mu_{x}, S_2(x_{adv},x) \geq 1-\epsilon_2^2$
  \STATE \hspace{0.5cm} $y_{adv} \leftarrow \text{model.predict}(x_{adv})$
  \STATE \hspace{0.5cm} $\mathbf{If}$ {$\text{label} \neq y_{adv}$}
  \STATE \hspace{1cm} $\mathbf{return}$ $x_{adv}$
  \STATE \hspace{0.5cm} $\mathbf{else}$
  \STATE \hspace{1cm} $x = x_{adv}$
  \STATE \hspace{0.5cm}   \textbf{return} $x_{adv}$
    \end{algorithmic}
\end{algorithm}

\subsection{Approximate Solution -- Faster PGAP}
We formulate an equivalent optimization problem from \eqref{eq:opt_ssim_s2} by relaxing Constraint 1 and substituting it in Constraint 2, and converting it into a minimization problem. The solution to this problem is given by:

\begin{equation}
\label{eq:cf_min}
x_{adv} = 1(\mu_{x}) + k_{22} + (\sqrt{k_{21}}) \left(\frac{\nabla_{x} \mathcal{L}(w,x,y)}{||\nabla_{x} \mathcal{L}(w,x,y)||} \right),
\end{equation}

where $k_{21}$, $k_{22}$ are defined in \eqref{eq:c2} and $1$ in $1(\mu_{x})$ is all ones of size $x_{adv}$; thus providing a closed form solution to \eqref{eq:opt_ssim_s2_approx}. We refer the reader to Appendix \ref{sec:approximate_solution_appendix} for the intermediate steps.

Algorithm \ref{alg:alg_iter_CF} presents the steps of FPGAP (a fast approximate variant of PGAP) by incorporating the closed-form solution above in the iterations.

\begin{algorithm}[ht]
\caption{Faster adversarial example generation}
\label{alg:alg_iter_CF}
  \begin{algorithmic}
  \STATE
  \STATE {\textsc{FPGAP}}($x, \text{model}, \text{label},\epsilon_2, \text{iterNum}$)
  \STATE \hspace{0.5cm} $i = 0$
  \STATE \hspace{0.5cm}  $\mathbf{While}$ {$i \leq \text{iterNum}$} $\mathbf{do}$
  \STATE \hspace{0.5cm} $\displaystyle x_{adv} \leftarrow 1(\mu_{x}) + k_{22} + (\sqrt{k_{21}}) \left(\frac{\nabla_{x} \mathcal{L}(w,x,y)}{||\nabla_{x} \mathcal{L}(w,x,y)||} \right)$
  \STATE \hspace{0.5cm} $y_{adv} \leftarrow \text{model.predict}(x_{adv})$
  \STATE \hspace{0.5cm} $\mathbf{If}$ {$\text{label} \neq y_{adv}$}
  \STATE \hspace{1cm} $\mathbf{return}$ $x_{adv}$
  \STATE \hspace{0.5cm} $\mathbf{else}$
  \STATE \hspace{1cm} $x = x_{adv}$
  \STATE \hspace{0.5cm}   \textbf{return} $x_{adv}$
    \end{algorithmic}
\end{algorithm}
\subsection{Some remarks}
We would like to point out that the structure of the presented approximate \eqref{eq:cf_min} solution is very similar to adversarial examples generated by FGM \cite{dong2018boosting} attack, which generates $l_2$ norm bounded additive perturbations. The expression for adversarial examples given by FGM is

\begin{equation}
\label{eq:fgm}
\begin{aligned}
x_{adv} = x + \epsilon \left(\frac{\nabla_{x} \mathcal{L}(w,x,y)}{||\nabla_{x} \mathcal{L}(w,x,y)||} \right).
\end{aligned}
\end{equation}

Both the formulations \eqref{eq:cf_min} and \eqref{eq:fgm} use normalized gradient \(\nabla_{x} \mathcal{L}(w,x,y)/||\nabla_{x} \mathcal{L}(w,x,y)||\). However the significant difference in the performance of the proposed attack is due to the presence of the terms $k_{21}$ and $k_{22}$ which are derived from the structure term in the SSIM index. It is very intriguing that such a simple change results in a huge improvement in the precision scores (Section \ref{sec:results_quant}) and fooling rate (Section \ref{sec:results_add}).

An alternate approach to find adversarial examples with high SSIM could be to subtract a differentiable version of SSIM from the loss function, similar to other such approaches in literature \cite{parimala2019quality}. Note that the existing norm bounded approaches impose a norm
ball constraint on the perturbation. This allows for the intermediate formulation (as in equation \ref{eq:formulation})
to be interpreted as a dual norm, hence enabling a closed-form solution. This would not be possible if the norm is
added directly to the loss function instead. Our formulation follows
a similar philosophy, where we impose a perceptual similarity
constraint instead of subtracting the SSIM from the loss function. Specifically, the mathematically amenable
variant of SSIM allows us to derive a closed-form solution
in this perceptually guided setting as well. As in the $l_p$ norm case, adding the SSIM index directly to the loss term does not admit a closed-form solution. In addition, the resulting non-convex formulation may also lead to increased
complexity and convergence-related issues.

\begin{figure*}[ht]
   \centering
\begin{tabular}{ccccc}

{\bf{PGAP}} (0.99) &SSIM map channel 1& SSIM map channel 2&SSIM map channel 3&GradCAM++ map original\\
\includegraphics[height = 0.75in,width = 0.9in]{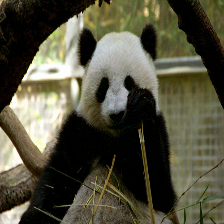}&
\includegraphics[height = 0.75in,width = 0.9in]{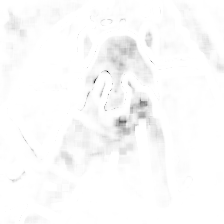}&
\includegraphics[height = 0.75in,width = 0.9in]{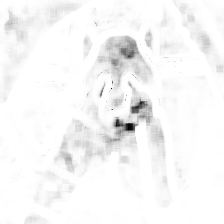}&
\includegraphics[height = 0.75in,width = 0.9in]{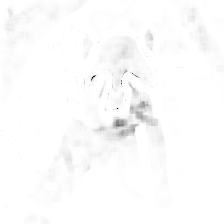}&
\includegraphics[height = 0.75in,width = 0.9in]{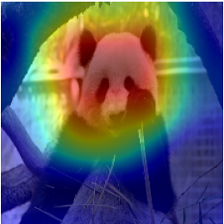}\\
{\bf{FPGAP}} (0.99) &&&&\\
\includegraphics[height = 0.75in,width = 0.9in]{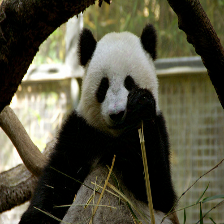}&
\includegraphics[height = 0.75in,width = 0.9in]{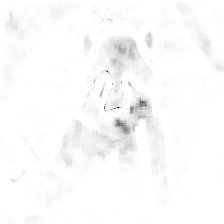}&
\includegraphics[height = 0.75in,width = 0.9in]{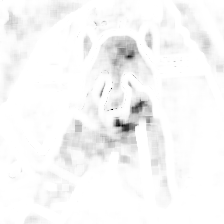}&
\includegraphics[height = 0.75in,width = 0.9in]{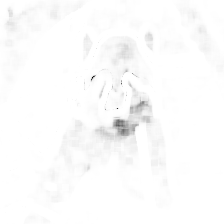}&
\includegraphics[height = 0.75in,width = 0.9in]{sup/panda_2//4_gradCAMPlus.PNG}\\
FGSM (0.44) &&&&\\
\includegraphics[height = 0.75in,width = 0.9in]{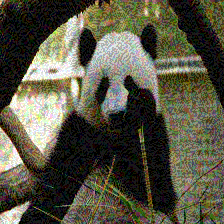}&
\includegraphics[height = 0.75in,width = 0.9in]{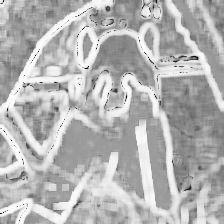}&
\includegraphics[height = 0.75in,width = 0.9in]{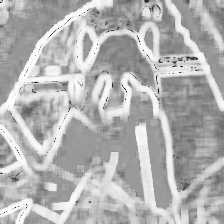}&
\includegraphics[height = 0.75in,width = 0.9in]{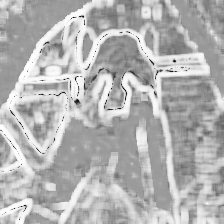}&
\includegraphics[height = 0.75in,width = 0.9in]{sup/panda_2//4_gradCAMPlus.PNG}\\
FGM\_ $l_2$ (0.82) &&&&\\
\includegraphics[height = 0.75in,width = 0.9in]{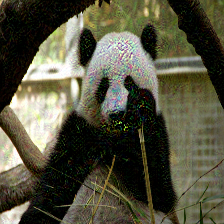}&
\includegraphics[height = 0.75in,width = 0.9in]{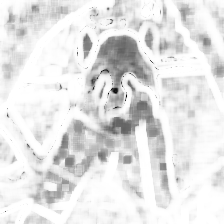}&
\includegraphics[height = 0.75in,width = 0.9in]{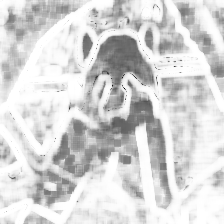}&
\includegraphics[height = 0.75in,width = 0.9in]{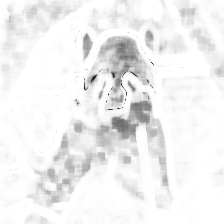}&
\includegraphics[height = 0.75in,width = 0.9in]{sup/panda_2//4_gradCAMPlus.PNG}\\
PGD (0.99) &&&&\\
\includegraphics[height = 0.75in,width = 0.9in]{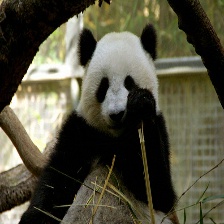}&
\includegraphics[height = 0.75in,width = 0.9in]{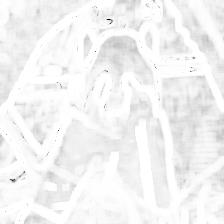}&
\includegraphics[height = 0.75in,width = 0.9in]{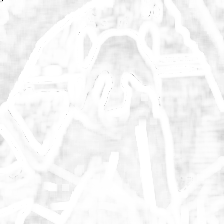}&
\includegraphics[height = 0.75in,width = 0.9in]{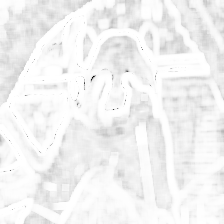}&
\includegraphics[height = 0.75in,width = 0.9in]{sup/panda_2//4_gradCAMPlus.PNG}\\
PGD\_ $l_2$ (0.99) &&&&\\
\includegraphics[height = 0.75in,width = 0.9in]{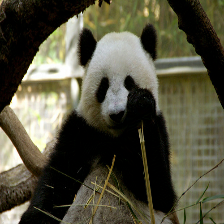}&
\includegraphics[height = 0.75in,width = 0.9in]{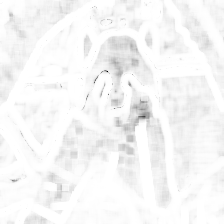}&
\includegraphics[height = 0.75in,width = 0.9in]{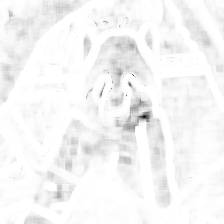}&
\includegraphics[height = 0.75in,width = 0.9in]{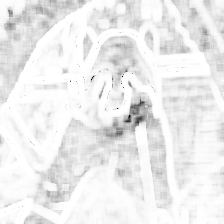}&
\includegraphics[height = 0.75in,width = 0.9in]{sup/panda_2//4_gradCAMPlus.PNG}\\
\small (a)&
\small (b)&
\small (c)&
\small (d)&
\small (e)

\end{tabular}
\caption{SSIM maps comparison of adversarial examples generated. (a): Adversarial perturbations with different methods and SSIM index value (rounded off to two decimal places),  (b),(c) and (d): SSIM maps of RGB channels respectively, (e): GradCAM++ output of original image.}
\label{fig:panda_full} 
\end{figure*}

\section{Results and Discussion}
\label{sec:results}
As discussed earlier, our key goal is to analyze the image regions which the adversarial perturbations affect the most. In particular, our interest is more on understanding the impact of adversarial perturbations on the structurally (perceptually) important regions of the image. We first start with a qualitative analysis with some illustrative examples and then quantify our observations using some well-known metrics. 

\begin{enumerate}
    \item \emph{Model used:} We use the MobileNetV2 \cite{sandler2018mobilenetv2} architecture with pre-trained weights, which has $72\%$ top-1 accuracy on the ImageNet validation set \cite{ILSVRC15} for quantitative and qualitative analysis. All the algorithms (PGAP, FPGAP, and the competing methods) are applied to this model to generate adversarial perturbations, one for each image in the ImageNet validation set.
    \item \emph{Methods being compared:} We compare our methods (PGAP, FPGAP) with other popular norm-bounded techniques-FGSM ($l_\infty$ bounded perturbations) \cite{goodfellow2014explaining}, FGM ($l_2$ bounded perturbations) \cite{dong2018boosting}, and PGD ($l_\infty$ and $l_2$ bounded perturbations) \cite{madry2017towards}.
\item \emph{Techniques used: }
    For each of the methods (FPGAP, FGSM, etc.), we generate the \emph{SSIM maps} between the adversarial examples and original images. As discussed in section \ref{sec:ssim_index}, SSIM maps \cite{wang2004image} provides a visualization of the distortions in the test image with respect to the reference image. Lighter regions in these maps correspond to lower distortion, while darker regions correspond to higher distortion levels. 


In order to obtain the semantically important regions of an image, we use Class Activation Maps. In particular, we use GradCAM++ \cite{chattopadhay2018grad}. The heat maps generated by GradCAM++ provide spatial regions that are most important for classification.
\item \emph{Hyper-parameter tuning:} Considering that each of the involved algorithms has different hyperparameters, we propose the following setup to compare them on an even ground. Consider the following Fooling Rate (FR) score: 

 \[\frac{1}{N}\sum_{i=1}^{N} \mathbbm{1}\left(\text{pre-attack label} (i)\neq \text{post-attack label}(i)\right), \]
where $N$ is the number of images in the dataset.
For a fair comparison, we select the parameters for different methods such that the Fooling Rate is $\approx 1$.

\end{enumerate}
\subsection{Qualitative Analysis}
\label{sec:results_qual}
\begin{figure*}
   \centering
   \subfloat[]{\includegraphics[width=2.5in]{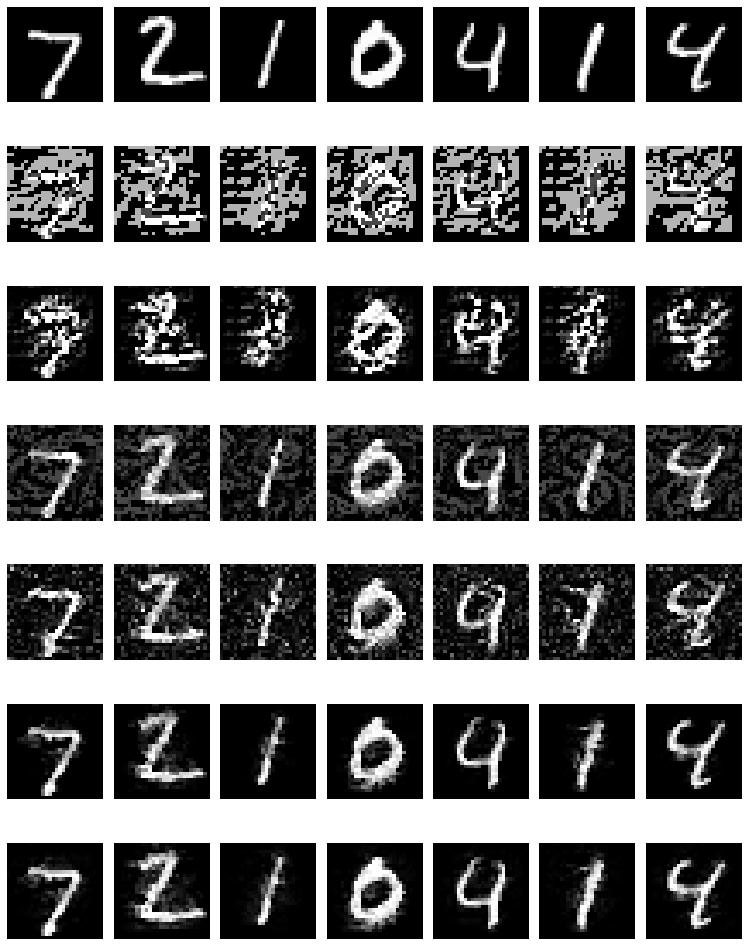}}%
\hfil
\subfloat[]{\includegraphics[width=2.5in]{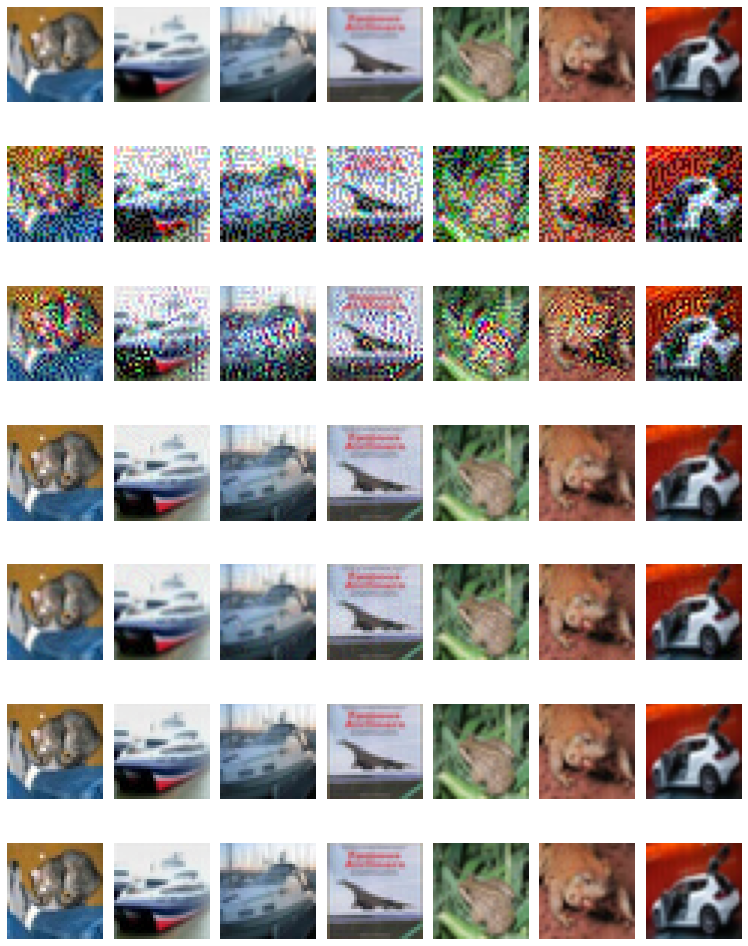}}%
\caption{Adversarial examples generated using multiple methods with similar fooling rate on images from the MNIST (left panel) and CIFAR-10 (right panel) datasets. From top to bottom each row corresponds to test images, FGSM, FGM\_$l2$, PGD, PGD\_$l2$, PGAP and FPGAP respectively.}
 \label{fig:imageComparision}
\end{figure*}

We first provide some illustrative examples that compare the SSIM maps generated by all the methods with GradCAM++ heat maps. For instance, consider Figure \ref{fig:panda_full}, we see that locations of the distortion in the proposed methods (PGAP, FPGAP) agree very well with the GradCAM++ maps. In particular, the number of distortions generated \emph{outside} the GradCAM++ regions of interest is very few. This, in turn provides evidence that our method does indeed perturb the regions in the image important for classification decisions. We can also observe (from the much the lighter maps corresponding to the proposed methods) that the proposed methods introduce far lesser perceptual distortions (or are more localized) compared to other methods.
    
Similar observation can be made from Figure \ref{fig:imageComparision} that illustrates performance on the MNIST and CIFAR-10 datasets. On MNIST data, the proposed method introduces perturbations in and around the digit present in the image, whereas the perturbations added by the other methods are distributed across the image. On the natural images in the CIFAR-10 dataset, our method is able to carefully identify regions important for classification decisions and introduces perturbations only in such regions. 
    
\subsection{Quantitative Analysis}
\label{sec:results_quant}
\begin{figure*}[t]
   \centering
\begin{tabular}{cccc}
\includegraphics[width = 3in,height = 1.75in]{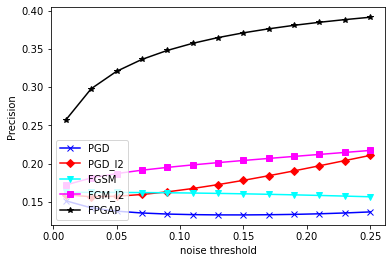}&
\includegraphics[width = 3in,height = 1.75in]{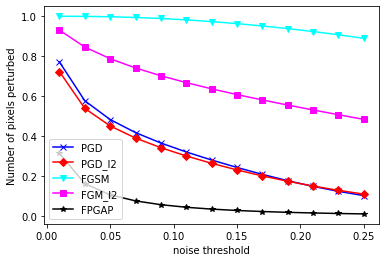}&\\
\includegraphics[width = 3in,height = 1.75in]{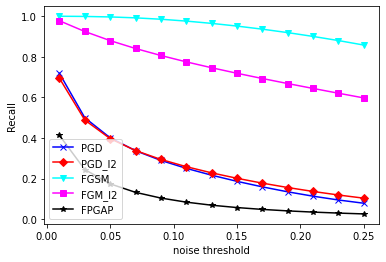}&
\includegraphics[width = 3in,height = 1.75in]{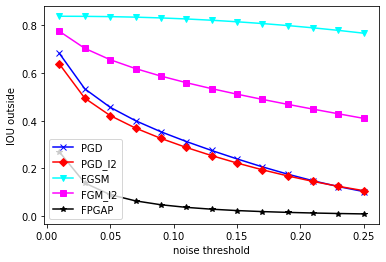}\\

\end{tabular}
\caption{Quantitative evaluation of adversarial perturbations using IOU based metrics (Avg precision, Avg number of pixels, Avg recall, Avg IOU outside) on ImageNet dataset.}

\label{fig:iou_metrics} 
\end{figure*}

In this section, we systematically quantify the observations made in the previous section. For this, we consider the following maps
\begin{itemize}
   \item \emph{Noise map corresponding the adversarial examples:} For
each image $i$ in the dataset, we generate a binary image $N_i$ given by
   \[
    N_i = \left\{(1-\text{SSIM map}(i))>\texttt{noise threshold} \right\},
    \]
  we evaluate each of the algorithms being compared at various values of \texttt{noise threshold}. 
    \item \emph{Noise map corresponding to GradCAM++:} For each image $i$ in the dataset, we generate the binary image $G_i$ given by 
    \[
    G_i = \left\{\text{GradCAM++ output}(i))>0.75\right\}.
    \]
\end{itemize} 
The binary image $N_i$ gives the regions affected by the adversarial perturbations, and the binary image $G_i$ gives the regions that are important for classification. We would like to note here that the GradCAM++ scale is normalized to lie between 0 and 1, with regions contributing to the classification decision taking on higher values. We empirically fixed the threshold on GradCAM++ to 0.75 to extract these important regions for classification (for example, the regions that are highlighted red in GradCAM++ in Figure \ref{fig:panda_full}). 

By considering the entries in $G_i$ to be the ground truth, we can use traditional metrics such as precision and IOU (intersection over union). We first compute the True Positives ($\texttt{TP}_i$), this is the number of positive pixels in $N_i$ that are also positive in $G_i$, and the number of false positives ($\texttt{FP}_i$), which is the number of pixels positive in $N_i$ but not in $G_i$. Similarly, we compute the true negatives ($\texttt{TN}_i$) and the false negatives ($\texttt{FN}_i$).
\begin{align*}
    &\texttt{TP}_i = \text{sum}(N_i\land G_i) & \texttt{FP}_i = \text{sum}(N_i\land \lnot G_i)\\
    &\texttt{FN}_i = \text{sum}(\lnot N_i \land G_i) &\texttt{TN}_i = \text{sum}(\lnot N_i \land \lnot G_i).
\end{align*}
Based on these, we evaluate the algorithms on the following metrics
\begin{enumerate}
    \item \emph{Precision:} the ratio of number of pixels perturbed in the important regions to the total number of perturbed pixels.
    \[
    \text{Precision}= \frac{1}{N}\sum_i \texttt{TP}_i/(\texttt{TP}_i + \texttt{FP}_i),
    \]
    where $N$ is the total number of images in the dataset. High values of this metric implies that the perturbations are precise in targeting the regions that are sensitive for classification. 
    \item \emph{Number of pixels perturbed:} This is the fraction of the number of positives in $N_i$, averaged over the dataset. 
\item \emph{Recall} ($= \sum_i\texttt{TP}_i/N(\texttt{TP}_i+\texttt{FN}_i)$ gives the ratio of number of pixels perturbed in the important regions to the total number of pixels in the important regions. Note that this metric may not be relevant enough for our setup: as an example, consider a trivial case where the perturbation perturbs all the pixels; in this case, the recall will be high, even though the perturbation is completely agnostic to the structure of the image. Nevertheless, we include this metric for completion. 
    \item \emph{Intersection over union (IOU)} Similar to other such metrics for image segmentation we can use
    \[
    \text{IOU}= \frac{1}{N}\sum_i \frac{\texttt{TP}_i}{\texttt{FP}_i+\texttt{TP}_i+\texttt{FN}_i} = \frac{1}{N}\sum_i \text{IOU}(N_i,G_i). 
    \]
    This is also the ratio of the intersection of the regions of $N_i$ and $G_i$ to their union. This metric has similar issues to recall, discussed above. To over come this, we define \emph{Intersection over union outside (IOU outside)}. Consider the following:
    \begin{align*}
    \text{IOU outside}&= \frac{1}{N}\sum_i \texttt{FP}_i/(\texttt{TP}_i+\texttt{FP}_i+\texttt{TN}_i) \\&= \frac{1}{N}\sum_i \text{IOU}(N_i,\lnot G_i). 
    \end{align*}
    This gives us the amount of perturbation the adversarial example generation method introduces in the regions that are not important for classification (i.e. outside $G_i$) averaged on entire data set. Ideally, if a perturbation only targets the important regions ($G_i$), the value of IOU outside should be very small.

\end{enumerate}

For each of the algorithms being compared, these scores are calculated for many different values of \texttt{noise threshold} and the resulting plots are shown in Fig \ref{fig:iou_metrics}. For reference, Table \ref{table: iou_1} lists the scores at a particular \texttt{noise threshold}.

\begin{table*}[htp]
\caption{Quantitative evaluation of adversarial perturbations using IOU based metrics on ImageNet dataset at noise threshold = 0.11.}
\begin{center}
\resizebox{0.75\textwidth}{!}{%
\begin{tabular}{|l||c|c|c|c|c|c|c|}
\hline
Attack ($\epsilon$) & Precision  & Number of pixels perturbed & Recall & IOU outside  \\
\hline
FGSM (0.5) & 0.1613 & 0.9817 & 0.9759 & 0.8263  \\
\hline
FGM\_$l2$ (100) & 0.1982 & 0.6682 & 0.7750 & 0.5594 \\
\hline
PGD (0.03) & 0.1329 & 0.3217 &0.2503 & 0.3128 \\
\hline
PGD\_$l2$ (5.0) & 0.1672 & 0.3017 & 0.2585 & 0.2880 \\
\hline
{\it{FPGAP}} (0.005) & 0.3577 & 0.0454 &0.0837 & 0.0369 \\
\hline
\end{tabular}
}
\end{center}
\label{table: iou_1}
\end{table*}
These experiments clearly validate the qualitative observations made earlier in section \ref{sec:results_qual}. Our method (FPGAP) beats other existing norm bounded methods by a factor of $\approx 2$ on the precision scores. The number of pixels perturbed by our method is also lower compared with the other methods. For FPGAP, the recall scores are low; this observation is in line with our earlier observations that FPGAP introduces a smaller number of perturbations (as compared to the FGSM, which introduces perturbations all over the image).  
\subsection{Additional Results}
\label{sec:results_add}
\begin{figure*}[t]
   \centering
\begin{tabular}{cccc}
MNIST & CIFAR-10 & ImageNet\\
\includegraphics[width = 2in]{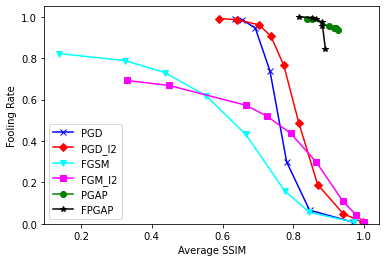}& 
\includegraphics[width = 2in]{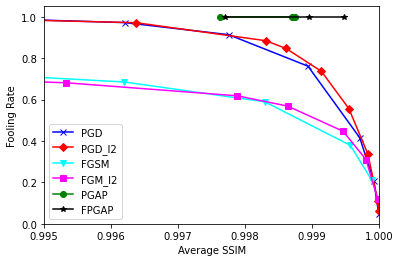}&
\includegraphics[width = 2in]{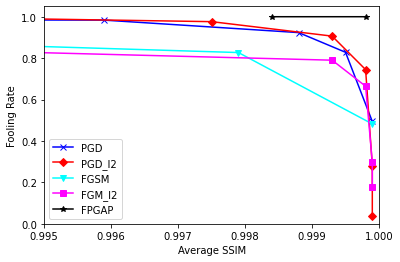}\\
\includegraphics[width = 2in]{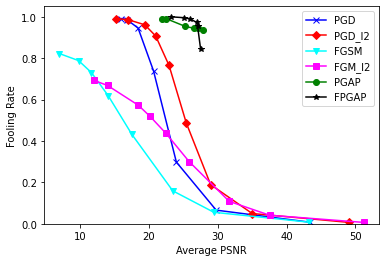}&
\includegraphics[width = 2in]{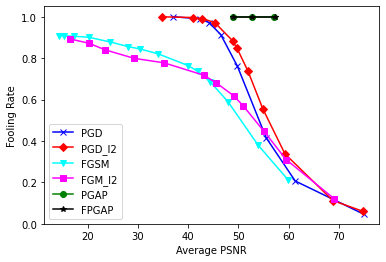}&
\includegraphics[width = 2in]{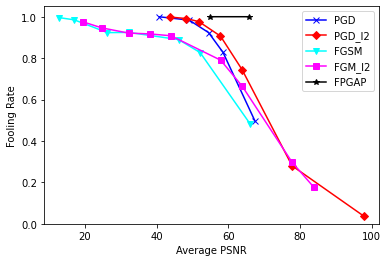}\\

 

\end{tabular}
\caption{Perceptual quality Average SSIM (top row) and Average PSNR (bottom row) versus Fooling Rate (FR) comparison of multiple adversarial perturbations on MNIST, CIFAR-10 and ImageNet datasets.}

\label{fig:qualFR} 
\end{figure*}

\begin{table*}[ht]
\caption{Fooling Rate comparison of different attacks on different datasets at a fixed quality \\ a) MNIST dataset (Average SSIM $\approx$ 0.85), b) CIFAR-10 dataset (Average SSIM $\approx$ 0.998), and c) ImageNet dataset (Average SSIM $\approx$ 0.998). Proposed methods are in italics.}
\centering
\begin{tabular}{|l||c|c|c|c|c|c|}
\hline
Attack & FGSM & FGM\_$l2$ & PGD & PGD\_$l2$  & {\it{FPGAP}}\\
\hline
Fooling Rate -- MNIST &  0.0556  & 0.3004 & 0.066 & 0.1891 &  0.9953\\
\hline
Fooling Rate -- CIFAR-10 & 0.5892  & 0.5681 & 0.7612 & 0.8472 &  1.0 \\
\hline
Fooling Rate -- ImageNet &  0.82696  & 0.79008 & 0.92332 & 0.97614 & 1.0\\
\hline
\end{tabular}
\label{table: ssim_all}
\end{table*}

\begin{table*}[ht]
\caption{Fooling Rate comparison of different attacks on different datasets at a fixed quality \\ a) MNIST dataset (Average PSNR $\approx$ 25), b) CIFAR-10 dataset (Average PSNR $\approx$  57), and c) ImageNet dataset (Average PSNR $\approx$ 65). Proposed methods are in italics.}
\centering
\begin{tabular}{|l||c|c|c|c|c|c|}
\hline
Attack & FGSM & FGM\_$l2$ & PGD & PGD\_$l2$ &   {\it{FPGAP}}\\
\hline
Fooling Rate -- MNIST &  0.1592  & 0.3004 & 0.2982 & 0.4867 &  0.9953\\
\hline
Fooling Rate -- CIFAR-10 & 0.2102 & 0.3099 &  0.4157 & 0.3376 & 1.0 \\
\hline
Fooling Rate -- ImageNet & 0.4822 & 0.66434 & 0.4942 & 0.74458 & 0.99998 \\
\hline
\end{tabular}
\label{table: psnr_all}
\end{table*}

In addition to the quantitative and qualitative analysis presented earlier, in this section, we analyze the perceptual quality of the adversarial examples generated by different methods. For each of the algorithms considered, we compute the structural similarity (SSIM) and PSNR between the adversarial examples and original images, averaged over the dataset. We plot the obtained Average SSIM and Average PSNR values vs FR (as defined in Section \ref{sec:results_qual}) in Figure \ref{fig:qualFR}.

From Figure \ref{fig:qualFR}, it is clear that  the approximation method FPGAP (Algorithm \ref{alg:alg_iter_CF}) does a fairly good job in approximating PGAP (Algorithm \ref{alg:alg_iter_AE}) on MNIST and CIFAR-10 datasets. Note that on the ImageNet dataset, we only report the results of FPGAP and ignore PGAP since it is computationally expensive to solve the QCQP iteratively. 

From Figure \ref{fig:qualFR}, we can see that the adversarial perturbations generated by the proposed method lead to a much higher FR (than norm bounded approaches) in the high-SSIM regime; thus, our technique is able to generate adversarial examples with minimal impact on the image quality across all the three datasets. For illustration, we consider a parameter configuration that leads to roughly the same Average SSIM for all the techniques (see Table \ref{table: ssim_all}). It is evident that the proposed method clearly achieves high FR within the given perceptual quality range. We also corroborate this using FR comparison at the same Average PSNR (see Table \ref{table: psnr_all}).

\subsection{Remarks}

For some data points, we observed that the QCQP was not solved accurately (the cvxpy logs showed a much higher duality gap), resulting in anomalous adversarial examples with a much lower SSIM. This, in turn, underestimates the true average SSIM of PGAP, leading to higher fooling rates at lower (estimated) SSIM. This underestimation can be observed in Figure \ref{fig:qualFR} (first column), where we see that the FR achieved by FPGAP is higher than the FR achieved by PGAP for the same average SSIM. This shortcoming is overcome by our Faster PGAP (FPGAP) solution.

As mentioned earlier, we find it very intriguing that incorporating an SSIM constraint (which enforces perceptual similarity) also makes the perturbations semantically meaningful. One potential hypothesis is that the constraints derived from the SSIM index restrict the space of perturbations, and together with the guidance provided by the gradients calculated from the network, enable us to find the semantically important regions while inducing minimal perturbations in those regions.

\section{Conclusion}
\label{sec:conclusion}
In this work, we investigated adversarial examples and their relationship to semantically significant regions of the image. Also, as a byproduct of trying to answer this question, we proposed a perceptually guided technique to generate adversarial examples that are structurally similar to the original image. By leveraging useful mathematical properties of the SSIM index, we presented a convex formulation to find adversarial examples. To the best of our knowledge, this is the first convex formulation that explicitly incorporates the SSIM index into the adversarial framework. In addition, we also provide a (fast) closed form approximation that enables solving the proposed convex formulation on large datasets. In fact, none of the existing adversarial example techniques that use image quality metrics have a closed-form solution. This is in stark contrast to other norm bounded techniques (PGD, FGSM, etc.), which employ a closed-form solution to generate adversarial perturbations.Our formulation also does not assume any model on the adversarial perturbations. We analyzed the adversarial perturbations generated by the proposed technique on images from the ImageNet validation set using SSIM maps and GradCAM++. By comparing the precision and IOU scores, we observed that, unlike standard techniques, the proposed technique is semantically-aware, i.e., it specifically targets the regions of the image that are important for classification. The proposed method also generates high-quality adversarial examples while achieving a Fooling Rate similar to comparable techniques. 

\appendix
\subsection{Constraints Analysis}
\label{sec:constraints_appendix}
We analyse the constraints of \eqref{eq:opt_ssim_s1_s2} in more detail in this section.

\begin{enumerate}
    \item \emph{Constraint 1:} The first constraint $S_1(x,x_{adv}) \geq 1-\epsilon_1^2 $ is a linear constraint (this forces $x_{adv}$ to lie in an intersection of two half-spaces). We can see this using \eqref{eq:s_1_2}  and \eqref{eq:nmse}:
 \[
||\mu_{{x}_{adv}}-\eta_1\mu_{x}||^2 \leq ||\mu_{x}||_2^2 (\eta_1)^2 + c_1 \left(\epsilon_1^2 \eta_1\right),
\]
or equivalently 
\begin{equation}
\label{eq:c1}
N(-\sqrt{k_{11}} + k_{12}) \leq 1^Tx_{adv}  \leq (\sqrt{k_{11}}+ k_{12})N,
\end{equation}
where
\begin{gather*}
k_{11} = (\eta_1)^2 ||\mu_{x}||_2^2  + c_1 \left(\epsilon_1^2\eta_1\right), \\
    k_{12} = \eta_1 \mu_{x},  \text{ and }  \eta_1 =1/(1-\epsilon_1^2).
\end{gather*}

\item \emph{Constraint 2:} The second constraint in \eqref{eq:opt_ssim_s1_s2} $S_2(x,x_{adv}) \geq 1-\epsilon_2^2$ is a quadratic constraint (this forces $x_{adv}$ to be in a high dimensional sphere). We can see this using \eqref{eq:s_1_2}:
\begin{align*}
&||(x_{adv}-1\mu_{{x}_{adv}})-\eta_2(x-1\mu_{x})||^2\\ &\quad \leq ||x-1\mu_{x}||_2^2 (\eta_2)^2 + c_2 (\epsilon_2^2\eta_2),
\end{align*}
or equivalently
\begin{equation}
\label{eq:c2}
||(x_{adv}-1\mu_{{x}_{adv}}) - k_{22}|| \leq \sqrt{k_{21}}; 
\end{equation}
 where
\begin{gather*}
    k_{21} =  ||x-1\mu_{x}||_2^2 (\eta_2)^2 + c_2 (\eta_2\epsilon_2^2),\\
    k_{22} = \eta_2 (x-1\mu_{x}) \text{ and } \eta_2 = 1/(1-\epsilon_2^2).
\end{gather*}
\end{enumerate}

Based on the constraints \eqref{eq:c1}, \eqref{eq:c2}, and the objective, the optimization problem in \eqref{eq:opt_ssim_s1_s2} is convex; in particular it is a Quadratically Constrained Quadratic Program (QCQP).

\subsection{Problem Approximation}
\label{sec:approximate_solution_appendix}
We formulate an equivalent optimization problem \eqref{eq:opt_ssim_s2_approx} from \eqref{eq:opt_ssim_s2}, by relaxing the first constraint and substituting it in the second. While this solution is an approximation to PGAP on account of relaxing one of the constraints, it has the advantage of having a closed form solution.

Consider the following approximation to \eqref{eq:opt_ssim_s2}.

\begin{equation}
\label{eq:opt_ssim_s2_approx}
\begin{aligned}
\argmin_{x_{adv}} \quad & -(x_{adv})^{T} \nabla_{x} \mathcal{L}(w,x,y) \\
\textrm{s.t.} \quad & ||(x_{adv}-1(\mu_{x})) - k_{22}||^2 \leq k_{21} 
\end{aligned}
\end{equation}

We first define the Lagrangian \cite{boyd2004convex} of the above problem:  
\begin{equation}
\label{eq:lagrangian_min}
\begin{split}
 L &= -(x_{adv})^{T} \nabla_{x} \mathcal{L}(w,x,y) \\
 & + \lambda (||(x_{adv}-1(\mu_{x})) - k_{22}||^2 - k_{21}) 
\end{split}
\end{equation}

We first take gradient of Lagrangian $L$ and equate it to zero:
\(\nabla_{x_{adv}} L = 0.\) This leads us to 
\begin{equation}
    \nabla_{x} \mathcal{L}(w,x,y) =2 \lambda  ((x_{adv}-1(\mu_{x})) - k_{22}).\label{eq:cf_1_min}
\end{equation}

Since the objective in \eqref{eq:opt_ssim_s2_approx} is linear, from geometrical intuition the maximum/minimum occurs only on the boundary of the sphere (specified by the constraint):

\[||(x_{adv}-1(\mu_{x})) - k_{22}||^2 = k_{21},\]
which can be rewritten using \eqref{eq:cf_1_min} as
\begin{equation}
\nabla_{x} \mathcal{L}(w,x,y)^T \nabla_{x} \mathcal{L}(w,x,y)  = 4\lambda^2 k_{21}
\label{eq:cf_2_min}
\end{equation}

Using the value of $\lambda$ from \eqref{eq:cf_2_min} in \eqref{eq:cf_1_min} we get:

\begin{equation*}
x_{adv} = 1(\mu_{x}) + k_{22} + (\sqrt{k_{21}}) \left(\frac{\nabla_{x} \mathcal{L}(w,x,y)}{||\nabla_{x} \mathcal{L}(w,x,y)||} \right),
\end{equation*}

where $k_{21}$, $k_{22}$ are defined in \eqref{eq:c2} and $1$ in $1(\mu_{x})$ is all ones of size $x_{adv}$; thus providing a closed form solution to \eqref{eq:opt_ssim_s2_approx}.

\subsection{Supplementary material}
In this work, we try to understand the landscape of adversarial perturbations through the perceptual quality lens. To achieve this goal, we rely on the perceptual quality metric SSIM index. We generate adversarial perturbations by maximizing the linear approximation of the loss function subject to the constraints derived from the mathematically amenable variant of the SSIM index. We use Carlini-Wagner (CW) loss function (with confidence parameter k =50) \cite{carlini2017towards} for generating adversarial examples. We analyze the perturbations generated by proposed method qualitatively and quantitatively and show the efficacy in terms of localization to semantically important regions compared to the norm-bounded adversarial perturbations. 


Tools used: Tensorflow \cite{45381}, Cvxpy \cite{diamond2016cvxpy}, Foolbox \cite{rauber2017foolbox}, tf-keras-vis \cite{keisentfkerasvis}.

\subsection{Qualitative Analysis}

To understand and analyze the impact of adversarial perturbations on the spatial regions of images, we use SSIM maps. Compared to the standard $l_p$ norm bounded perturbations, the proposed approach generates perturbations that are perceptually-aware (structure-aware) and able to find the regions that are important for classification. We provide corroborative evidence of the same using GradCAM++ output. 


\subsubsection{SSIM Map}
SSIM maps contain local SSIM index values at pixel level calculated using pixels in the local neighbourhood of 8$\times$8 block or Gaussian window of size 11$\times$11 centered at the pixel. The global SSIM index value is calculated by using/pooling these local SSIM index values. Typically these maps are defined for grayscale images or one colour plane of colour images, here we present the SSIM maps of three channels (R, G, and B). These maps help us to visualize the distortion locally at a pixel level. In this work, we use these maps to locate the image regions that are affected by adversarial perturbations. For example, in Figure \ref{fig:bagel_full} the dark pixels in the second column of images are the locations where adversarial perturbations are introduced in the channel one (Red), and columns three and four corresponds to channels two and three (Green and Blue).  

\subsubsection{Supporting Results}
To further illustrate efficacy of our method we present several examples (Figures: \ref{fig:bagel_full}, \ref{fig:perfume_full}, \ref{fig:golfball_full}, \ref{fig:dog_full}, \ref{fig:dragonfly_full}, \ref{fig:elephant_full} and \ref{fig:frog_full}) of the proposed method along with other norm bounded perturbations in the following:

We observed that the iterative methods (PGD, PGD$\_l2$, PGAP, FPGAP) are doing well compared to the basic (non-iterative) methods (FGSM and FGM). In iterative approaches, by updating the image at each iteration, non-linearity is introduced in crafting the adversarial perturbations, which could be the reason for better performance. To understand these iterative methods further, we analyse the impact of adversarial perturbations on spatial regions of images using absolute difference maps.

\subsubsection{Understanding Iterative Methods}
We use random test images from MNIST and CIFAR-10 test datasets for this analysis. To understand and analyze the impact of adversarial perturbations on the spatial regions of images, we use absolute difference maps. These maps are generated by taking the absolute difference of adversarial images with respect to original image.These maps will help us in locating the perturbations introduced in different parts of the image. Following is the setup we use for this analysis: 

\begin{enumerate}
    \item Fix number of iterations for all the methods (for example: 10, 30 etc).
    \item Find the smallest $\epsilon$ to generate the adversarial example (for random test images at fixed number of iterations specified in step 1).
    \item Compute the absolute difference map, PSNR and SSIM with respect to original image. 
\end{enumerate}

Figures \ref{fig:mnist_psnr_10},  \ref{fig:mnist_psnr_30}, \ref{fig:mnist_psnr_50} and \ref{fig:mnist_psnr_70} are generated using random test images from the MNIST test images (at different iterations: 10, 30, 50 and 70 respectively). We make the following observations from these figures:

\begin{enumerate}
    \item From the absolute difference maps we can say that the proposed method introduces perturbations only at important regions of the image (around the digit present in the image) in all iteration settings.
    \item Since the proposed method is able to identify the structural regions and adding perturbations only in those regions the required amount of perturbation is small compared to other methods which adds perturbations all over the image. We believe that this could be the reason for high PSNR and SSIM index values of adversarial images generated by proposed method compared to the other methods (PGD and PGD$\_l2$).
    \item As the number of iterations increase the performance of PGD and PGD$\_l2$ is improving. In particular, as we increase the number of iterations the perturbations introduced by PGD$\_l2$ are more localized around the digit area.
    \item The proposed method seems to achieve higher PSNR and SSIM index values in fewer iterations compared to other methods.
    
\end{enumerate}

Similar observations can be made on the CIFAR-10 test images (Figures \ref{fig:cifar_psnr_10},  \ref{fig:cifar_psnr_30}, \ref{fig:cifar_psnr_50} and \ref{fig:cifar_psnr_70}).

\subsection{Additional Results}
\begin{figure*}
   \centering
\begin{tabular}{cccc}

\includegraphics[width = 2in]{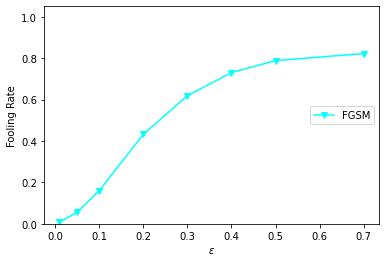}& 
\includegraphics[width = 2in]{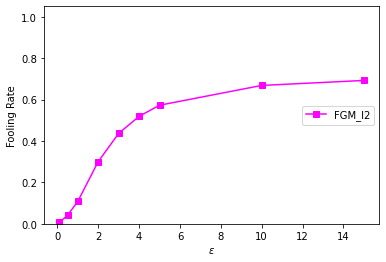}&
\includegraphics[width = 2in]{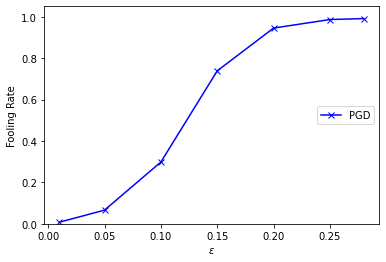}\\

\includegraphics[width = 2in]{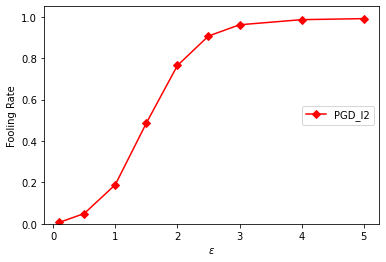}&
\includegraphics[width = 2in]{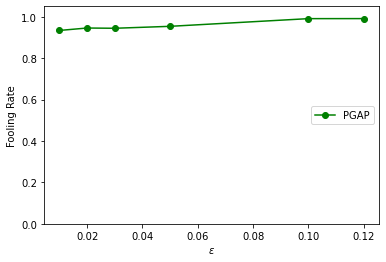}&
\includegraphics[width = 2in]{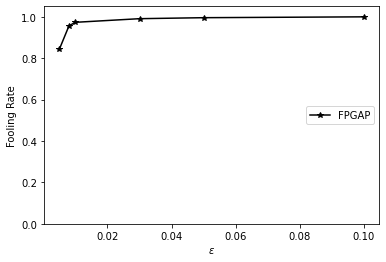}\\

\end{tabular}
\caption{$\epsilon$ versus Fooling Rate (FR) comparison of multiple adversarial perturbations on MNIST.}

\label{fig:eps_fr_mnist} 
\end{figure*}

\begin{figure*}
   \centering
\begin{tabular}{cccc}

\includegraphics[width = 2in]{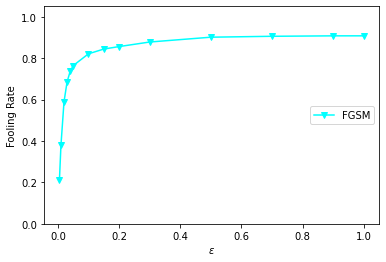}&
\includegraphics[width = 2in]{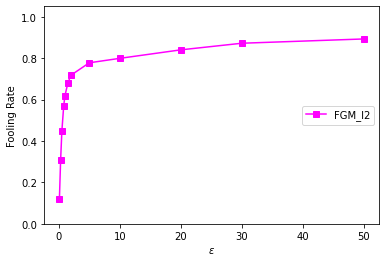}&
\includegraphics[width = 2in]{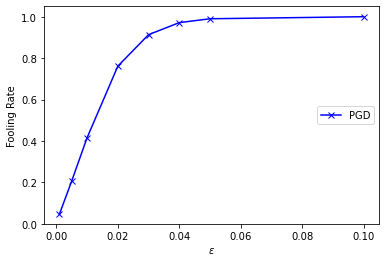}\\

\includegraphics[width = 2in]{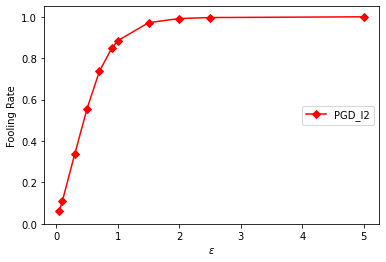}&
\includegraphics[width = 2in]{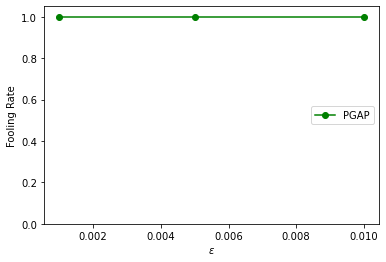}&
\includegraphics[width = 2in]{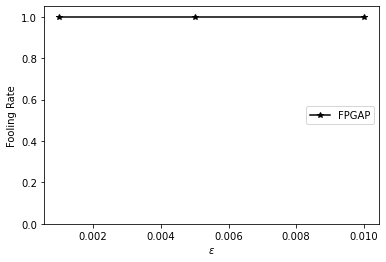}\\

\end{tabular}
\caption{$\epsilon$ versus Fooling Rate (FR) comparison of multiple adversarial perturbations on CIFAR-10.}

\label{fig:eps_fr_cifar} 
\end{figure*}

\begin{figure*}
   \centering
\begin{tabular}{cccc}

\includegraphics[width = 2in]{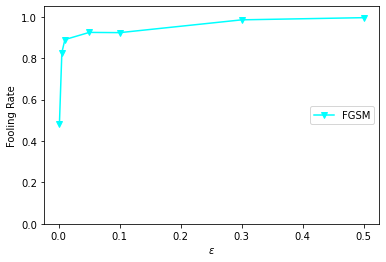}&
\includegraphics[width = 2in]{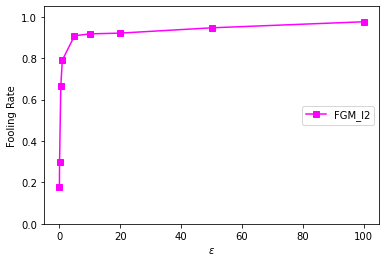}&
\includegraphics[width = 2in]{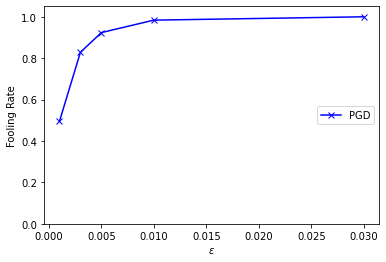}\\

\includegraphics[width = 2in]{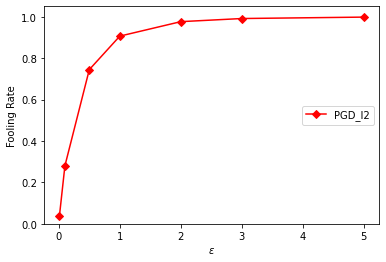}&
\includegraphics[width = 2in]{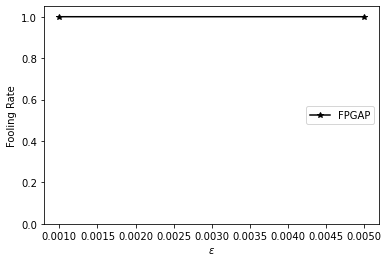}&\\

\end{tabular}
\caption{$\epsilon$ versus Fooling Rate (FR) comparison of multiple adversarial perturbations on ImageNet.}

\label{fig:eps_fr_imagenet} 
\end{figure*}

\begin{figure*}
   \centering
\begin{tabular}{ccccc}

{\bf{PGAP}} (1.00) &SSIM map channel 1& SSIM map channel 2&SSIM map channel 3&GradCAM++ map original\\
\includegraphics[width = 1in]{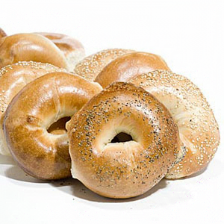}&
\includegraphics[width = 1in]{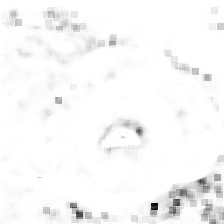}&
\includegraphics[width = 1in]{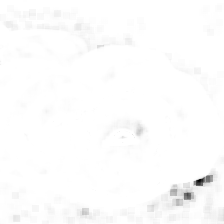}&
\includegraphics[width = 1in]{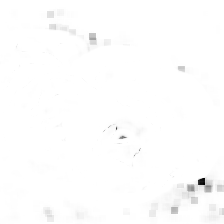}&
\includegraphics[width = 1in]{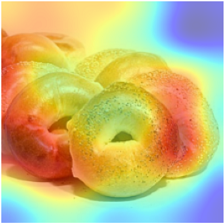}\\
{\bf{FPGAP}} (1.00) &&&&\\
\includegraphics[width = 1in]{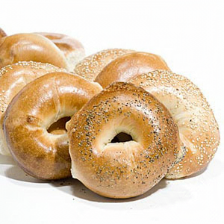}&
\includegraphics[width = 1in]{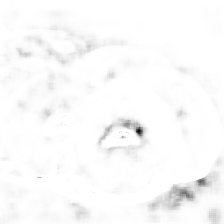}&
\includegraphics[width = 1in]{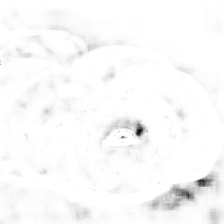}&
\includegraphics[width = 1in]{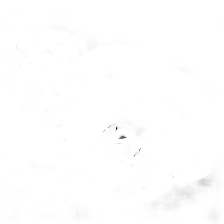}&
\includegraphics[width = 1in]{sup/bagel//11_gradCAMPlus.PNG}\\
FGSM (0.92) &&&&\\
\includegraphics[width = 1in]{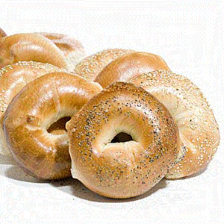}&
\includegraphics[width = 1in]{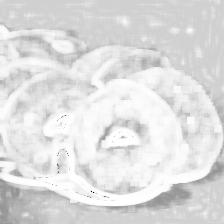}&
\includegraphics[width = 1in]{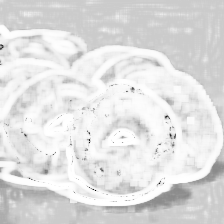}&
\includegraphics[width = 1in]{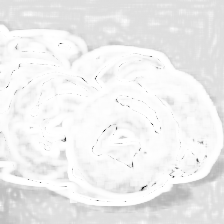}&
\includegraphics[width = 1in]{sup/bagel//11_gradCAMPlus.PNG}\\
FGM\_ $l_2$ (1.00) &&&&\\
\includegraphics[width = 1in]{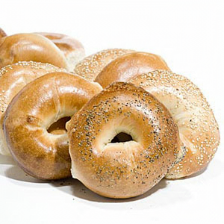}&
\includegraphics[width = 1in]{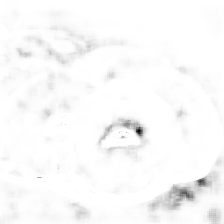}&
\includegraphics[width = 1in]{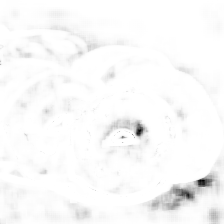}&
\includegraphics[width = 1in]{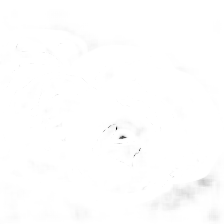}&
\includegraphics[width = 1in]{sup/bagel//11_gradCAMPlus.PNG}\\
PGD (1.00) &&&&\\
\includegraphics[width = 1in]{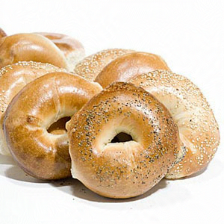}&
\includegraphics[width = 1in]{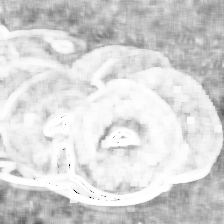}&
\includegraphics[width = 1in]{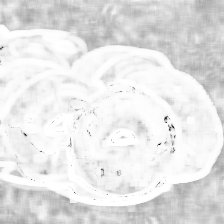}&
\includegraphics[width = 1in]{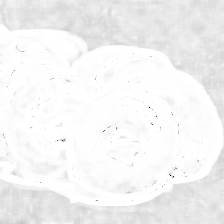}&
\includegraphics[width = 1in]{sup/bagel//11_gradCAMPlus.PNG}\\
PGD\_ $l_2$ (1.00) &&&&\\
\includegraphics[width = 1in]{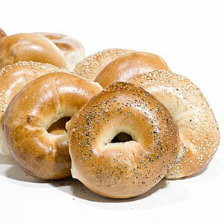}&
\includegraphics[width = 1in]{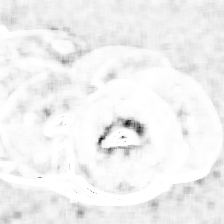}&
\includegraphics[width = 1in]{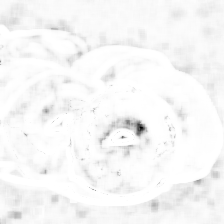}&
\includegraphics[width = 1in]{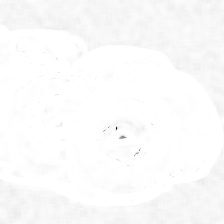}&
\includegraphics[width = 1in]{sup/bagel//11_gradCAMPlus.PNG}\\
\small (a)&
\small (b)&
\small (c)&
\small (d)&
\small (e)

\end{tabular}
\caption{SSIM maps comparison of adversarial examples generated. (a): Adversarial perturbations with different methods and SSIM index value (rounded off to two decimal places),  (b),(c) and (d): SSIM maps of RGB channels respectively, (e): GradCAM++ output of original image.}
\label{fig:bagel_full} 
\end{figure*}

\begin{figure*}
   \centering
\begin{tabular}{ccccc}

{\bf{PGAP}} (1.00) &SSIM map channel 1& SSIM map channel 2&SSIM map channel 3&GradCAM++ map original\\
\includegraphics[width = 1in]{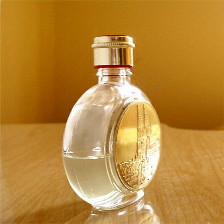}&
\includegraphics[width = 1in]{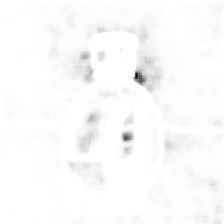}&
\includegraphics[width = 1in]{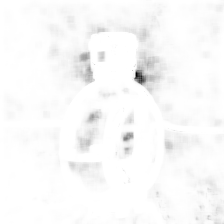}&
\includegraphics[width = 1in]{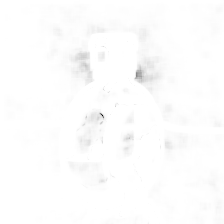}&
\includegraphics[width = 1in]{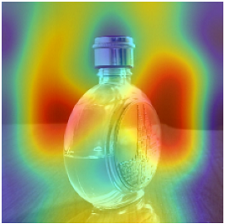}\\
{\bf{FPGAP}} (1.00) &&&&\\
\includegraphics[width = 1in]{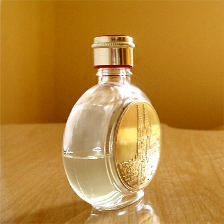}&
\includegraphics[width = 1in]{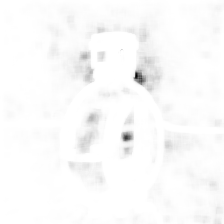}&
\includegraphics[width = 1in]{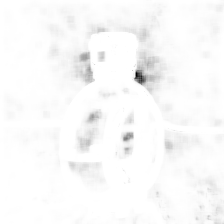}&
\includegraphics[width = 1in]{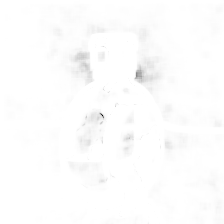}&
\includegraphics[width = 1in]{sup/perfume//36_gradCAMPlus.PNG}\\
FGSM (0.94) &&&&\\
\includegraphics[width = 1in]{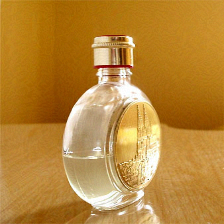}&
\includegraphics[width = 1in]{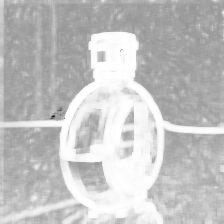}&
\includegraphics[width = 1in]{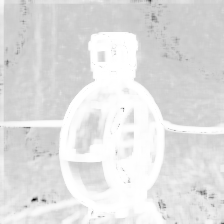}&
\includegraphics[width = 1in]{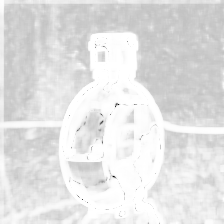}&
\includegraphics[width = 1in]{sup/perfume//36_gradCAMPlus.PNG}\\
FGM\_ $l_2$ (1.00) &&&&\\
\includegraphics[width = 1in]{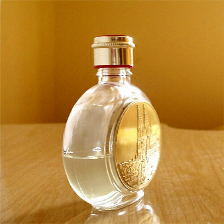}&
\includegraphics[width = 1in]{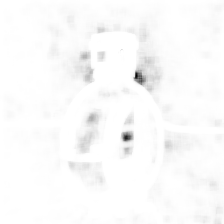}&
\includegraphics[width = 1in]{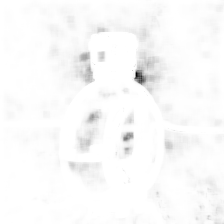}&
\includegraphics[width = 1in]{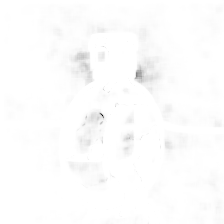}&
\includegraphics[width = 1in]{sup/perfume//36_gradCAMPlus.PNG}\\
PGD (0.99) &&&&\\
\includegraphics[width = 1in]{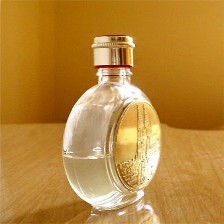}&
\includegraphics[width = 1in]{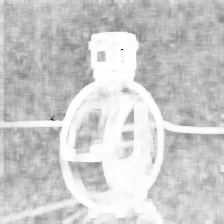}&
\includegraphics[width = 1in]{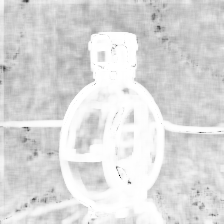}&
\includegraphics[width = 1in]{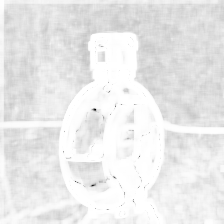}&
\includegraphics[width = 1in]{sup/perfume//36_gradCAMPlus.PNG}\\
PGD\_ $l_2$ (1.00) &&&&\\
\includegraphics[width = 1in]{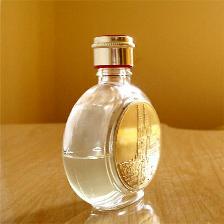}&
\includegraphics[width = 1in]{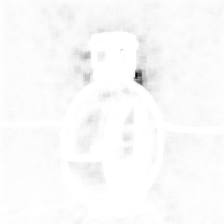}&
\includegraphics[width = 1in]{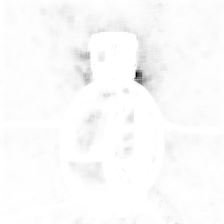}&
\includegraphics[width = 1in]{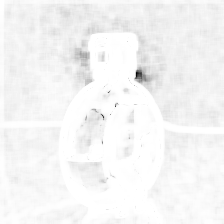}&
\includegraphics[width = 1in]{sup/perfume//36_gradCAMPlus.PNG}\\
\small (a)&
\small (b)&
\small (c)&
\small (d)&
\small (e)

\end{tabular}
\caption{SSIM maps comparison of adversarial examples generated. (a): Adversarial perturbations with different methods and SSIM index value (rounded off to two decimal places),  (b),(c) and (d): SSIM maps of RGB channels respectively, (e): GradCAM++ output of original image.}
\label{fig:perfume_full} 
\end{figure*}

\begin{figure*}
   \centering
\begin{tabular}{ccccc}

{\bf{PGAP}} (0.99) &SSIM map channel 1& SSIM map channel 2&SSIM map channel 3&GradCAM++ map original\\
\includegraphics[width = 1in]{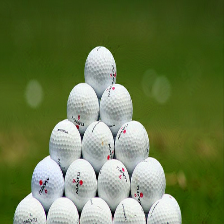}&
\includegraphics[width = 1in]{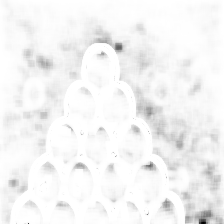}&
\includegraphics[width = 1in]{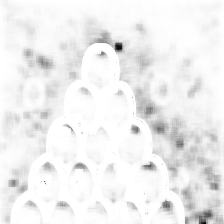}&
\includegraphics[width = 1in]{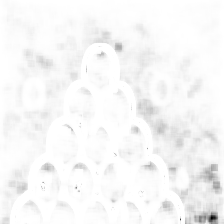}&
\includegraphics[width = 1in]{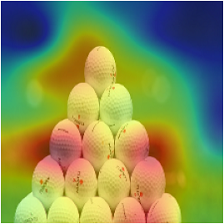}\\
{\bf{FPGAP}} (0.99) &&&&\\
\includegraphics[width = 1in]{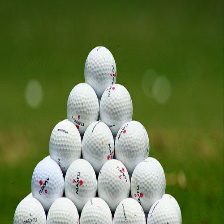}&
\includegraphics[width = 1in]{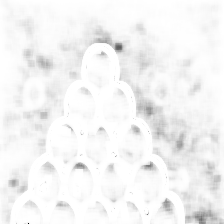}&
\includegraphics[width = 1in]{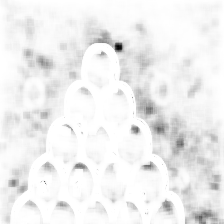}&
\includegraphics[width = 1in]{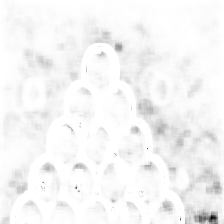}&
\includegraphics[width = 1in]{sup/golfball//31_gradCAMPlus.PNG}\\
FGSM (0.35) &&&&\\
\includegraphics[width = 1in]{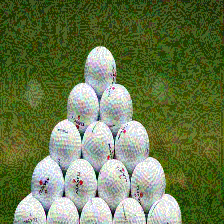}&
\includegraphics[width = 1in]{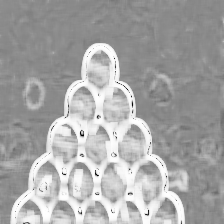}&
\includegraphics[width = 1in]{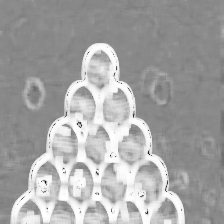}&
\includegraphics[width = 1in]{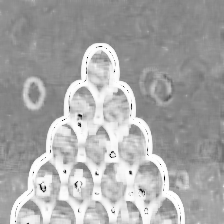}&
\includegraphics[width = 1in]{sup/golfball//31_gradCAMPlus.PNG}\\
FGM\_ $l_2$ (0.50) &&&&\\
\includegraphics[width = 1in]{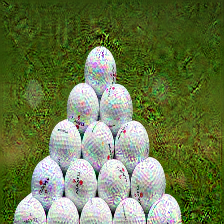}&
\includegraphics[width = 1in]{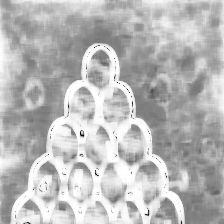}&
\includegraphics[width = 1in]{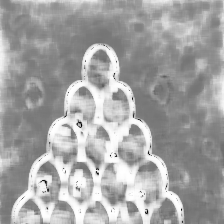}&
\includegraphics[width = 1in]{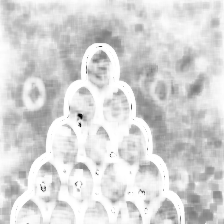}&
\includegraphics[width = 1in]{sup/golfball//31_gradCAMPlus.PNG}\\
PGD (0.99) &&&&\\
\includegraphics[width = 1in]{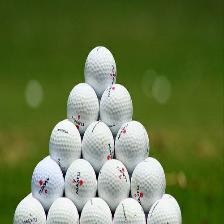}&
\includegraphics[width = 1in]{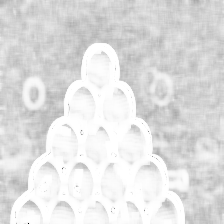}&
\includegraphics[width = 1in]{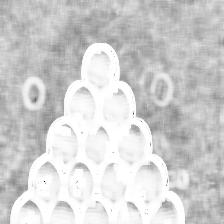}&
\includegraphics[width = 1in]{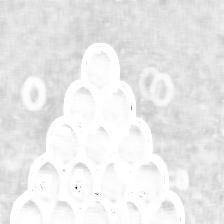}&
\includegraphics[width = 1in]{sup/golfball//31_gradCAMPlus.PNG}\\
PGD\_ $l_2$ (1.00) &&&&\\
\includegraphics[width = 1in]{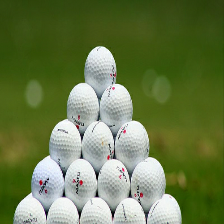}&
\includegraphics[width = 1in]{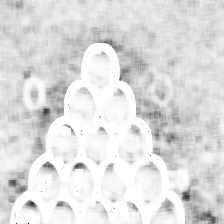}&
\includegraphics[width = 1in]{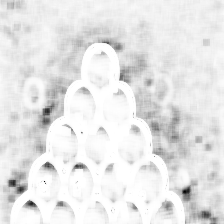}&
\includegraphics[width = 1in]{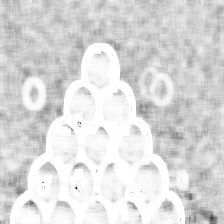}&
\includegraphics[width = 1in]{sup/golfball//31_gradCAMPlus.PNG}\\
\small (a)&
\small (b)&
\small (c)&
\small (d)&
\small (e)

\end{tabular}
\caption{SSIM maps comparison of adversarial examples generated. (a): Adversarial perturbations with different methods and SSIM index value (rounded off to two decimal places),  (b),(c) and (d): SSIM maps of RGB channels respectively, (e): GradCAM++ output of original image.}
\label{fig:golfball_full} 
\end{figure*}

\begin{figure*}
   \centering
\begin{tabular}{ccccc}

{\bf{PGAP}} (1.00) &SSIM map channel 1& SSIM map channel 2&SSIM map channel 3&GradCAM++ map original\\
\includegraphics[width = 1in]{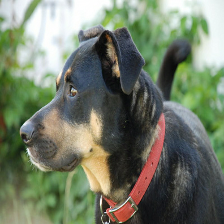}&
\includegraphics[width = 1in]{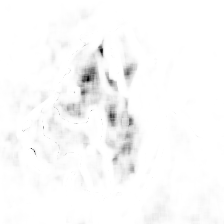}&
\includegraphics[width = 1in]{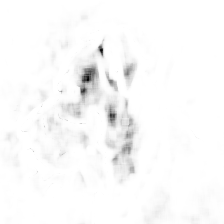}&
\includegraphics[width = 1in]{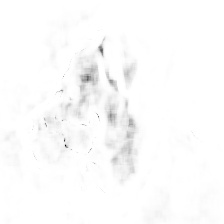}&
\includegraphics[width = 1in]{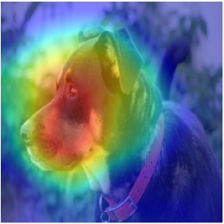}\\
{\bf{FPGAP}} (1.00) &&&&\\
\includegraphics[width = 1in]{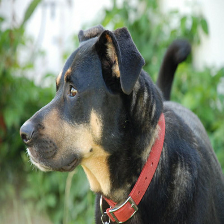}&
\includegraphics[width = 1in]{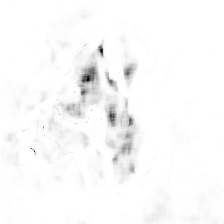}&
\includegraphics[width = 1in]{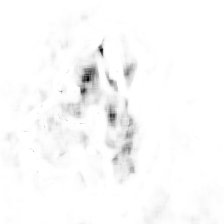}&
\includegraphics[width = 1in]{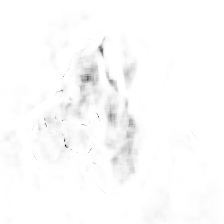}&
\includegraphics[width = 1in]{dog/1_gradCAMPlus.PNG}\\
FGSM (0.91) &&&&\\
\includegraphics[width = 1in]{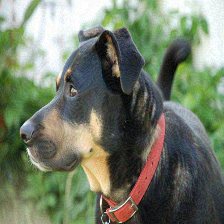}&
\includegraphics[width = 1in]{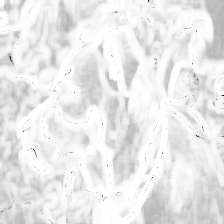}&
\includegraphics[width = 1in]{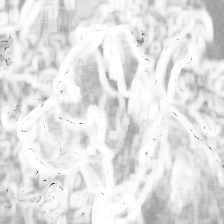}&
\includegraphics[width = 1in]{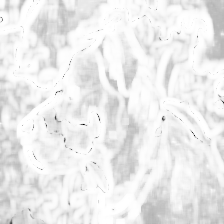}&
\includegraphics[width = 1in]{dog/1_gradCAMPlus.PNG}\\
FGM\_ $l_2$ (1.00) &&&&\\
\includegraphics[width = 1in]{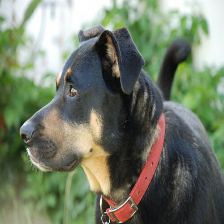}&
\includegraphics[width = 1in]{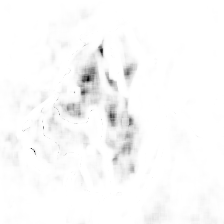}&
\includegraphics[width = 1in]{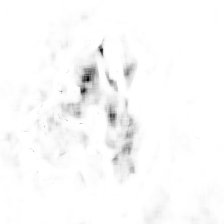}&
\includegraphics[width = 1in]{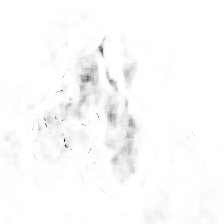}&
\includegraphics[width = 1in]{dog/1_gradCAMPlus.PNG}\\
PGD (1.00) &&&&\\
\includegraphics[width = 1in]{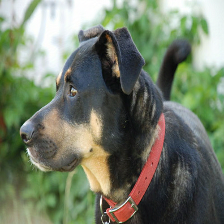}&
\includegraphics[width = 1in]{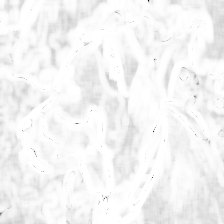}&
\includegraphics[width = 1in]{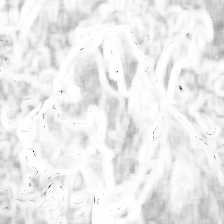}&
\includegraphics[width = 1in]{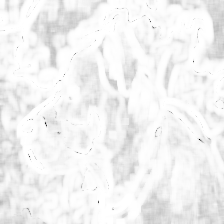}&
\includegraphics[width = 1in]{dog/1_gradCAMPlus.PNG}\\
PGD\_ $l_2$ (1.00) &&&&\\
\includegraphics[width = 1in]{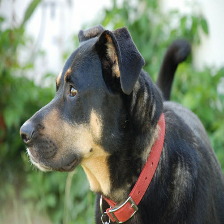}&
\includegraphics[width = 1in]{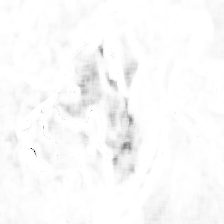}&
\includegraphics[width = 1in]{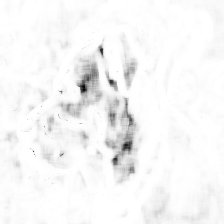}&
\includegraphics[width = 1in]{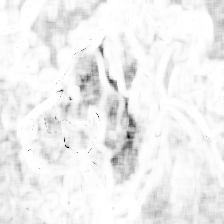}&
\includegraphics[width = 1in]{dog/1_gradCAMPlus.PNG}\\
\small (a)&
\small (b)&
\small (c)&
\small (d)&
\small (e)

\end{tabular}
\caption{SSIM maps comparison of adversarial examples generated. (a): Adversarial perturbations with different methods and SSIM index value (rounded off to two decimal places),  (b),(c) and (d): SSIM maps of RGB channels respectively, (e): GradCAM++ output of original image.}
\label{fig:dog_full} 
\end{figure*}

\begin{figure*}
   \centering
\begin{tabular}{ccccc}

{\bf{PGAP}} (1.00) &SSIM map channel 1& SSIM map channel 2&SSIM map channel 3&GradCAM++ map original\\
\includegraphics[width = 1in]{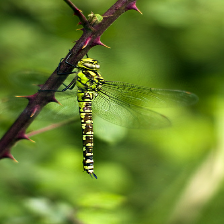}&
\includegraphics[width = 1in]{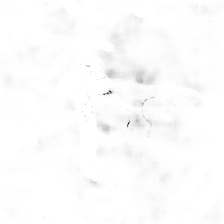}&
\includegraphics[width = 1in]{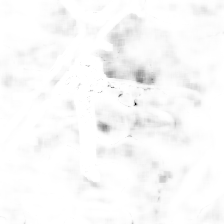}&
\includegraphics[width = 1in]{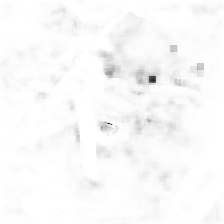}&
\includegraphics[width = 1in]{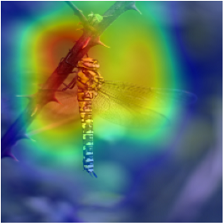}\\
{\bf{FPGAP}} (1.00) &&&&\\
\includegraphics[width = 1in]{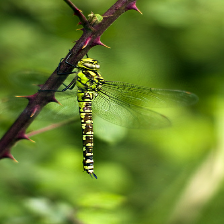}&
\includegraphics[width = 1in]{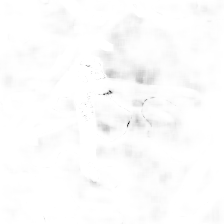}&
\includegraphics[width = 1in]{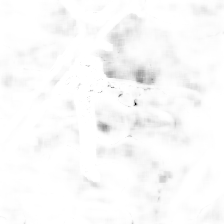}&
\includegraphics[width = 1in]{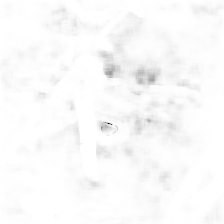}&
\includegraphics[width = 1in]{dragonfly/18_gradCAMPlus.PNG}\\
FGSM (0.20) &&&&\\
\includegraphics[width = 1in]{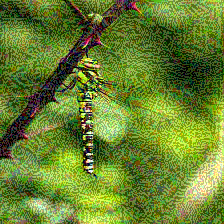}&
\includegraphics[width = 1in]{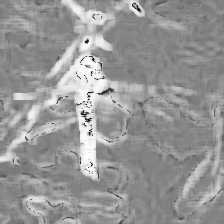}&
\includegraphics[width = 1in]{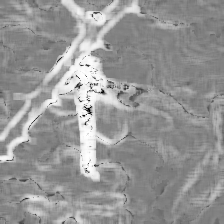}&
\includegraphics[width = 1in]{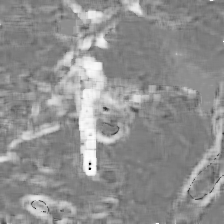}&
\includegraphics[width = 1in]{dragonfly/18_gradCAMPlus.PNG}\\
FGM\_ $l_2$ (0.48) &&&&\\
\includegraphics[width = 1in]{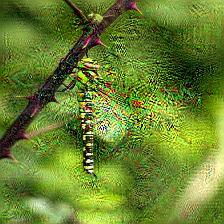}&
\includegraphics[width = 1in]{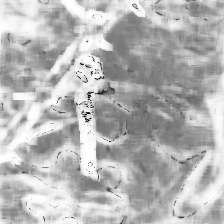}&
\includegraphics[width = 1in]{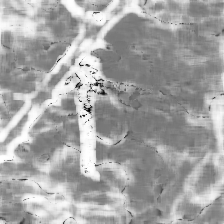}&
\includegraphics[width = 1in]{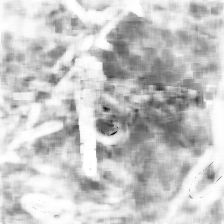}&
\includegraphics[width = 1in]{dragonfly/18_gradCAMPlus.PNG}\\
PGD (0.99) &&&&\\
\includegraphics[width = 1in]{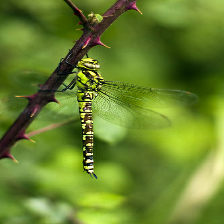}&
\includegraphics[width = 1in]{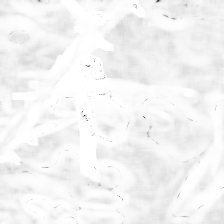}&
\includegraphics[width = 1in]{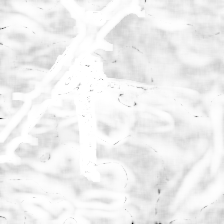}&
\includegraphics[width = 1in]{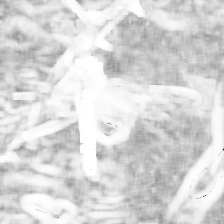}&
\includegraphics[width = 1in]{dragonfly/18_gradCAMPlus.PNG}\\
PGD\_ $l_2$ (1.00) &&&&\\
\includegraphics[width = 1in]{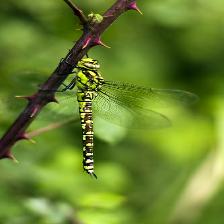}&
\includegraphics[width = 1in]{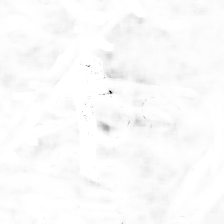}&
\includegraphics[width = 1in]{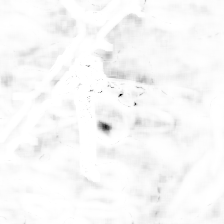}&
\includegraphics[width = 1in]{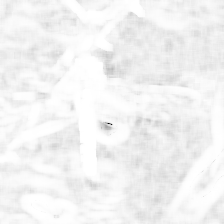}&
\includegraphics[width = 1in]{dragonfly/18_gradCAMPlus.PNG}\\
\small (a)&
\small (b)&
\small (c)&
\small (d)&
\small (e)

\end{tabular}
\caption{SSIM maps comparison of adversarial examples generated. (a): Adversarial perturbations with different methods and SSIM index value (rounded off to two decimal places),  (b),(c) and (d): SSIM maps of RGB channels respectively, (e): GradCAM++ output of original image.}
\label{fig:dragonfly_full} 
\end{figure*}

\begin{figure*}
   \centering
\begin{tabular}{ccccc}

{\bf{PGAP}} (1.00) &SSIM map channel 1& SSIM map channel 2&SSIM map channel 3&GradCAM++ map original\\
\includegraphics[width = 1in]{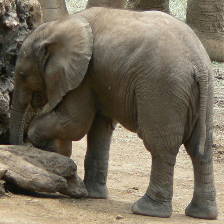}&
\includegraphics[width = 1in]{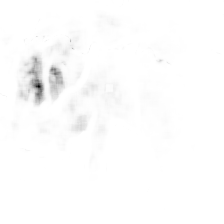}&
\includegraphics[width = 1in]{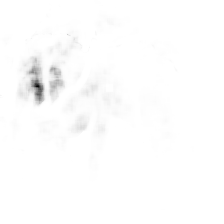}&
\includegraphics[width = 1in]{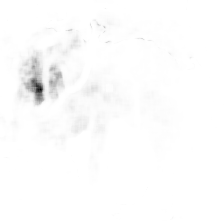}&
\includegraphics[width = 1in]{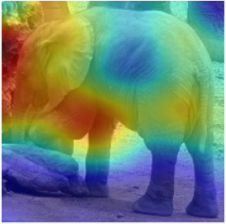}\\
{\bf{FPGAP}} (1.00) &&&&\\
\includegraphics[width = 1in]{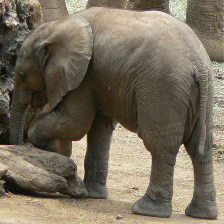}&
\includegraphics[width = 1in]{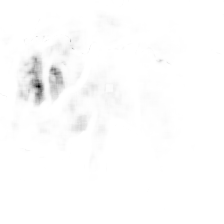}&
\includegraphics[width = 1in]{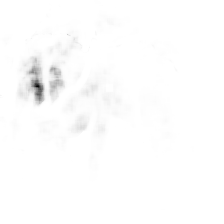}&
\includegraphics[width = 1in]{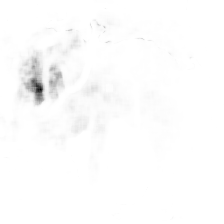}&
\includegraphics[width = 1in]{sup/elephant//26_gradCAMPlus.PNG}\\
FGSM (0.98) &&&&\\
\includegraphics[width = 1in]{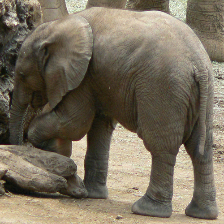}&
\includegraphics[width = 1in]{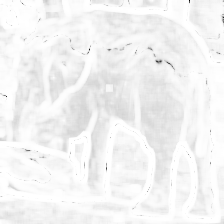}&
\includegraphics[width = 1in]{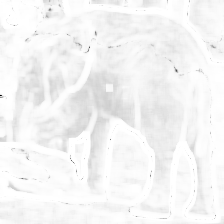}&
\includegraphics[width = 1in]{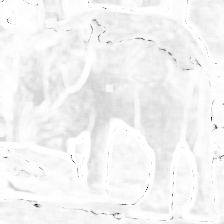}&
\includegraphics[width = 1in]{sup/elephant//26_gradCAMPlus.PNG}\\
FGM\_ $l_2$ (1.00) &&&&\\
\includegraphics[width = 1in]{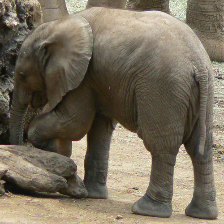}&
\includegraphics[width = 1in]{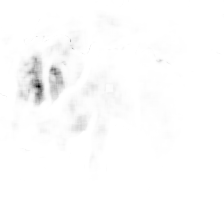}&
\includegraphics[width = 1in]{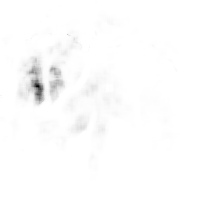}&
\includegraphics[width = 1in]{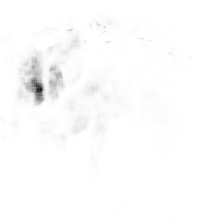}&
\includegraphics[width = 1in]{sup/elephant//26_gradCAMPlus.PNG}\\
PGD (1.00) &&&&\\
\includegraphics[width = 1in]{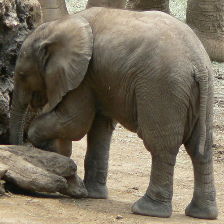}&
\includegraphics[width = 1in]{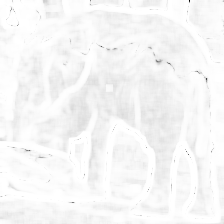}&
\includegraphics[width = 1in]{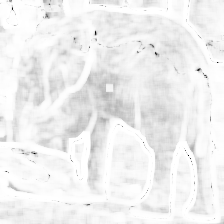}&
\includegraphics[width = 1in]{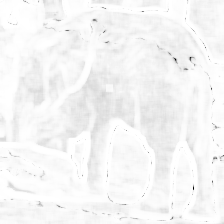}&
\includegraphics[width = 1in]{sup/elephant//26_gradCAMPlus.PNG}\\
PGD\_ $l_2$ (1.00) &&&&\\
\includegraphics[width = 1in]{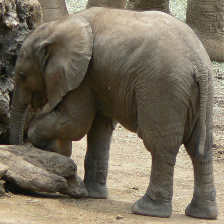}&
\includegraphics[width = 1in]{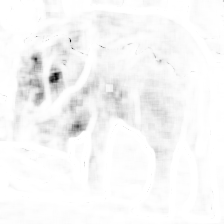}&
\includegraphics[width = 1in]{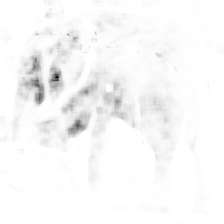}&
\includegraphics[width = 1in]{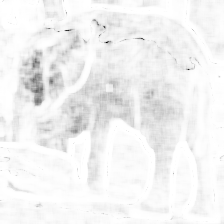}&
\includegraphics[width = 1in]{sup/elephant//26_gradCAMPlus.PNG}\\
\small (a)&
\small (b)&
\small (c)&
\small (d)&
\small (e)

\end{tabular}
\caption{SSIM maps comparison of adversarial examples generated. (a): Adversarial perturbations with different methods and SSIM index value (rounded off to two decimal places),  (b),(c) and (d): SSIM maps of RGB channels respectively, (e): GradCAM++ output of original image.}
\label{fig:elephant_full} 
\end{figure*}

\begin{figure*}
   \centering
\begin{tabular}{ccccc}

{\bf{PGAP}} (1.00) &SSIM map channel 1& SSIM map channel 2&SSIM map channel 3&GradCAM++ map original\\
\includegraphics[width = 1in]{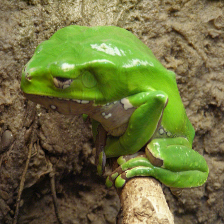}&
\includegraphics[width = 1in]{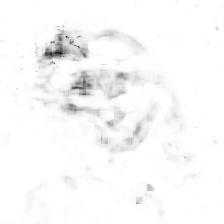}&
\includegraphics[width = 1in]{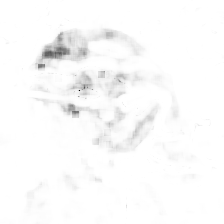}&
\includegraphics[width = 1in]{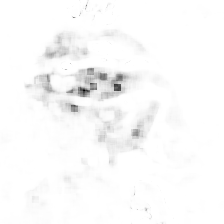}&
\includegraphics[width = 1in]{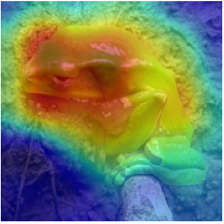}\\
{\bf{FPGAP}} (1.00) &&&&\\
\includegraphics[width = 1in]{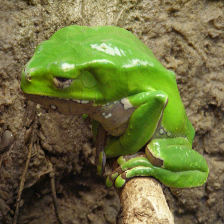}&
\includegraphics[width = 1in]{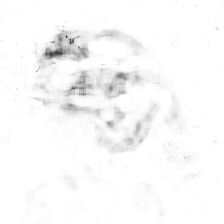}&
\includegraphics[width = 1in]{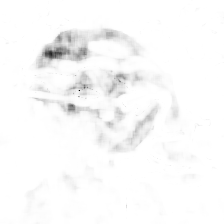}&
\includegraphics[width = 1in]{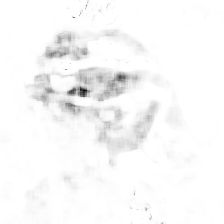}&
\includegraphics[width = 1in]{frog/13_gradCAMPlus.PNG}\\
FGSM (0.94) &&&&\\
\includegraphics[width = 1in]{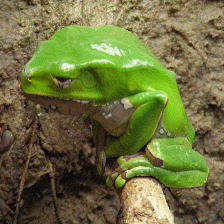}&
\includegraphics[width = 1in]{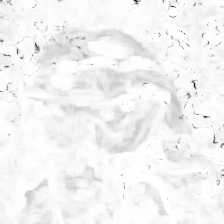}&
\includegraphics[width = 1in]{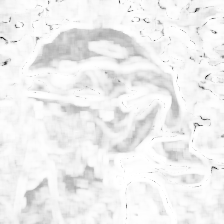}&
\includegraphics[width = 1in]{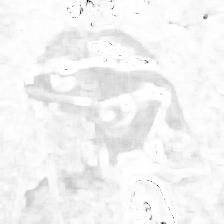}&
\includegraphics[width = 1in]{frog/13_gradCAMPlus.PNG}\\
FGM\_ $l_2$ (0.97) &&&&\\
\includegraphics[width = 1in]{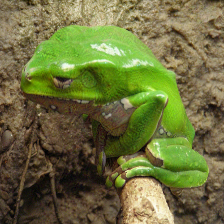}&
\includegraphics[width = 1in]{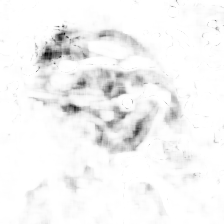}&
\includegraphics[width = 1in]{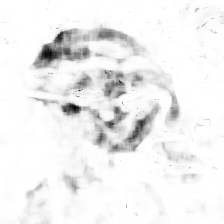}&
\includegraphics[width = 1in]{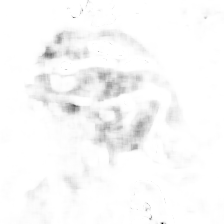}&
\includegraphics[width = 1in]{frog/13_gradCAMPlus.PNG}\\
PGD (1.00) &&&&\\
\includegraphics[width = 1in]{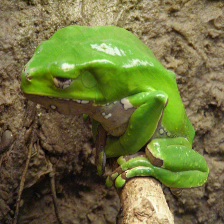}&
\includegraphics[width = 1in]{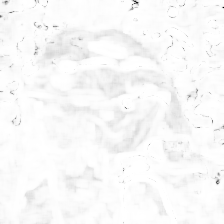}&
\includegraphics[width = 1in]{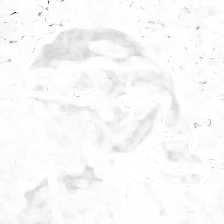}&
\includegraphics[width = 1in]{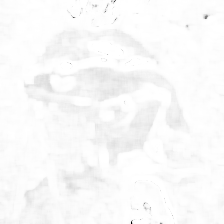}&
\includegraphics[width = 1in]{frog/13_gradCAMPlus.PNG}\\
PGD\_ $l_2$ (1.00) &&&&\\
\includegraphics[width = 1in]{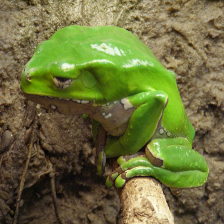}&
\includegraphics[width = 1in]{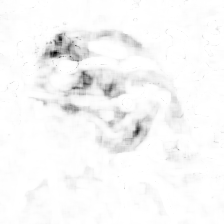}&
\includegraphics[width = 1in]{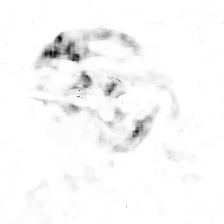}&
\includegraphics[width = 1in]{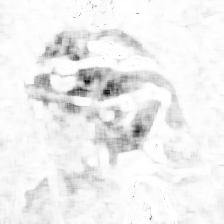}&
\includegraphics[width = 1in]{frog/13_gradCAMPlus.PNG}\\
\small (a)&
\small (b)&
\small (c)&
\small (d)&
\small (e)

\end{tabular}
\caption{SSIM maps comparison of adversarial examples generated. (a): Adversarial perturbations with different methods and SSIM index value (rounded off to two decimal places),  (b),(c) and (d): SSIM maps of RGB channels respectively, (e): GradCAM++ output of original image.}
\label{fig:frog_full} 
\end{figure*}

We compare our method with FGSM ($l_\infty$ bounded perturbations) \cite{goodfellow2014explaining}, FGM ($l_2$ bounded perturbations) \cite{dong2018boosting}, and PGD ($l_\infty$ and $l_2$ bounded perturbations) \cite{madry2017towards}. Note that for iterative methods the number of iterations used are 50. Each of the techniques under consideration has a parameter $\epsilon$ that controls the fooling rate: for norm bounded perturbations this is the largest allowed norm of the perturbation. For the proposed method, the parameter is $\epsilon_2$ ((9) in the main draft) which bounds the structural distortion in the generated adversarial example. We generate adversarial examples at different parameter values on MNIST, CIFAR-10 and ImageNet datasets for all these techniques. We can see the relation between the parameter $\epsilon$ and the Fooling Rate for these methods in Figures \ref{fig:eps_fr_mnist}, \ref{fig:eps_fr_cifar} and \ref{fig:eps_fr_imagenet}.     

However, the parameter $\epsilon$ has different role/significance (and different range of values) for each method. This makes it difficult to compare the success of different methods with respect to $\epsilon$. Hence we take a different route, and compare the imperceptibility and the Fooling Rate of the attack. To measure the imperceptibility we use image quality metric SSIM and PSNR. We compute Average SSIM and Average PSNR over the dataset at different parameter values (presented in Figures \ref{fig:eps_fr_mnist}, \ref{fig:eps_fr_cifar} and \ref{fig:eps_fr_imagenet}) for each method. Then we compare these with the Fooling Rate (Figure 6 in the main draft).

\begin{figure*}
   \centering
\begin{tabular}{ccc}
\includegraphics[height =3.5in, width = 2in]{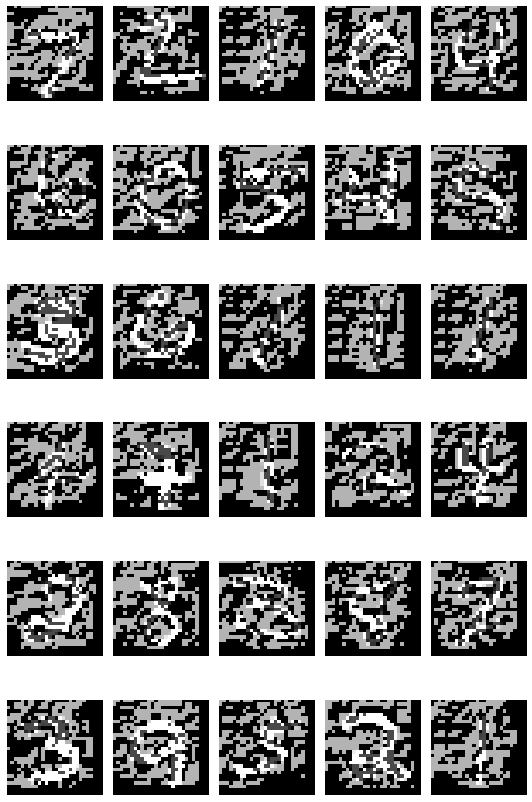}&
\includegraphics[height =3.5in, width = 2in]{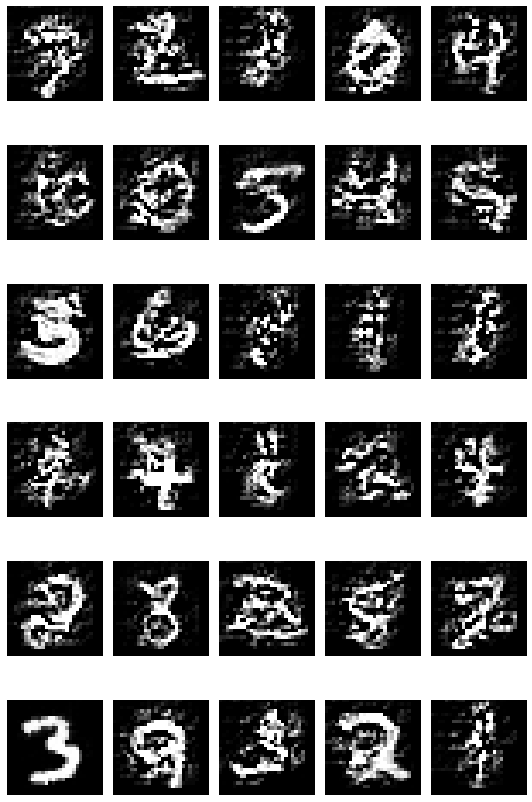}&
\includegraphics[height =3.5in, width = 2in]{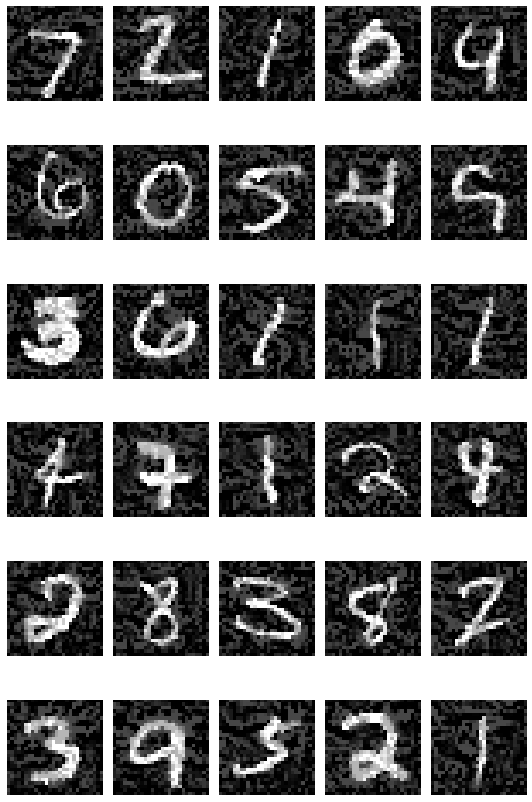}\\
FGSM & FGM\_$l2$ & PGD\\

\includegraphics[height =3.5in, width = 2in]{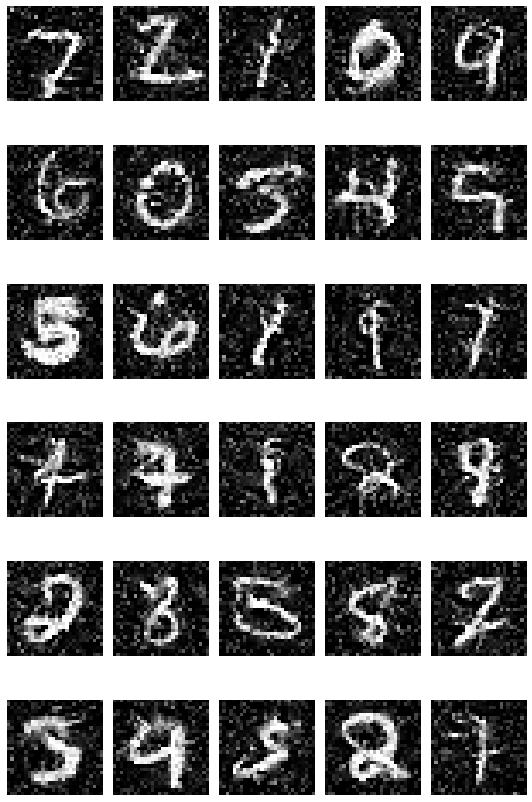}&
\includegraphics[height =3.5in, width = 2in]{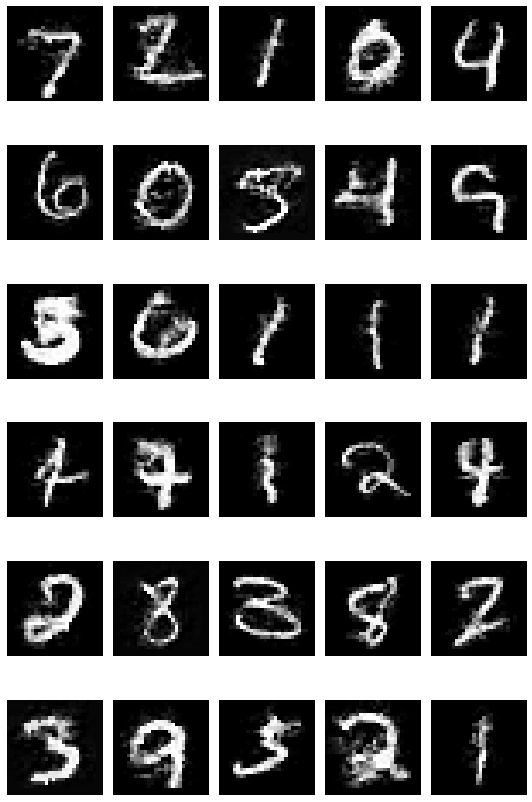}&
\includegraphics[height =3.5in, width = 2in]{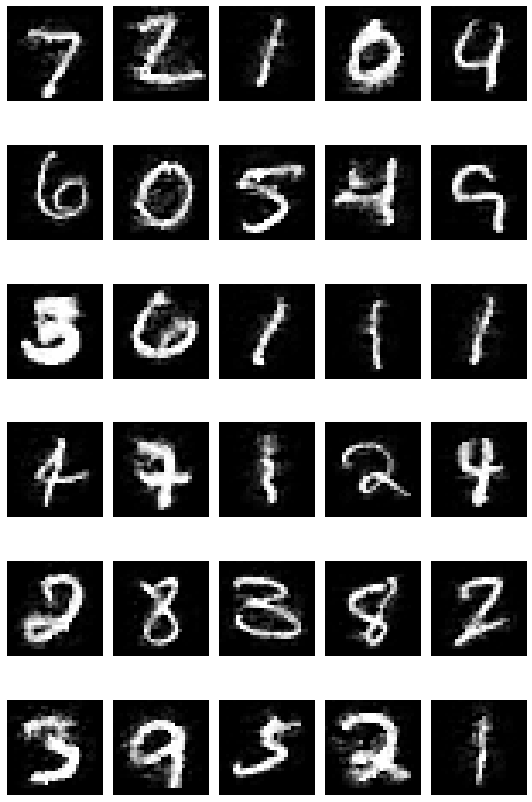}\\
PGD\_$l2$ & {\bf{PGAP}} & {\bf{FPGAP}} \\

\end{tabular}
\caption{Adversarial examples generated using multiple methods with similar fooling rate on MNIST dataset.}

\label{fig:mnist_all} 
\end{figure*}

\begin{figure*}
   \centering
\begin{tabular}{ccc}
\includegraphics[height =3.5in, width = 2in]{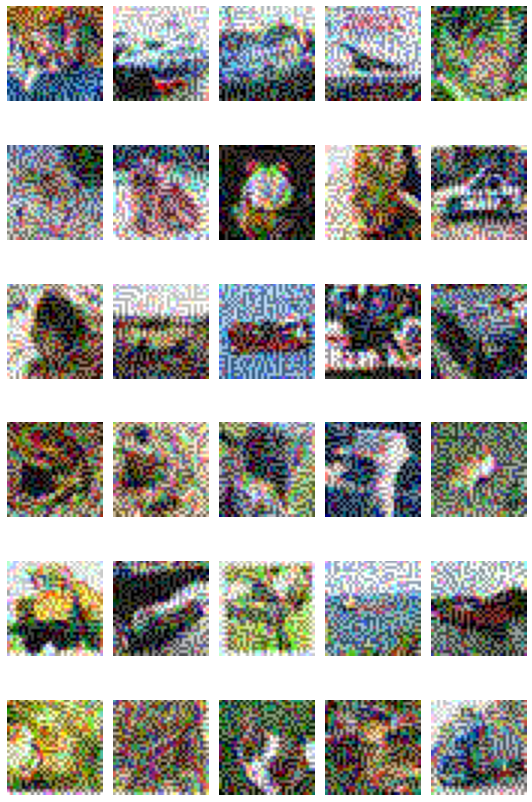}&
\includegraphics[height =3.5in, width = 2in]{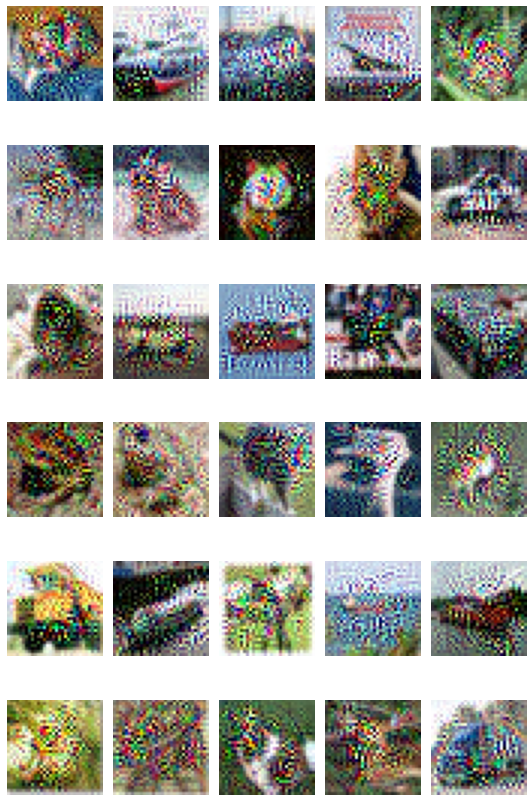}&
\includegraphics[height =3.5in, width = 2in]{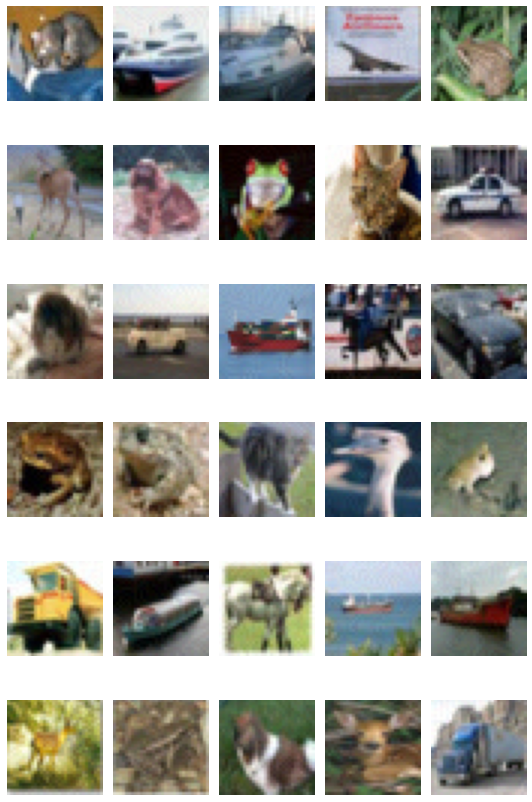}\\
FGSM & FGM\_$l2$ & PGD\\

\includegraphics[height =3.5in, width = 2in]{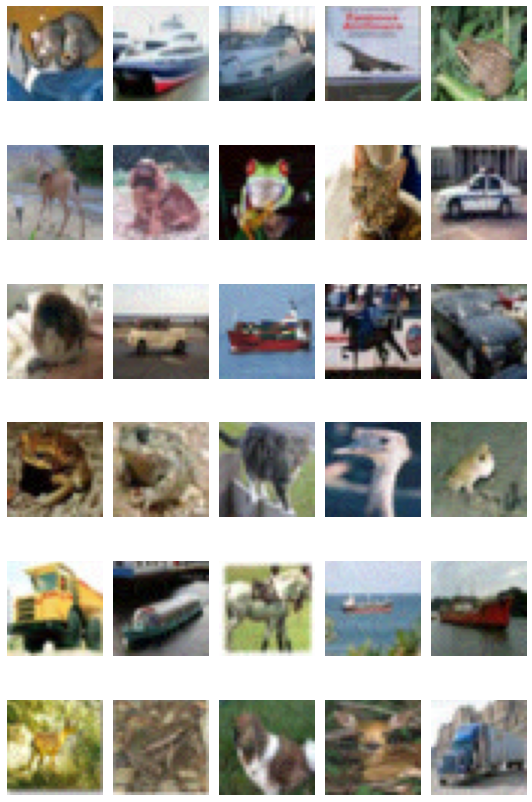}&
\includegraphics[height =3.5in, width = 2in]{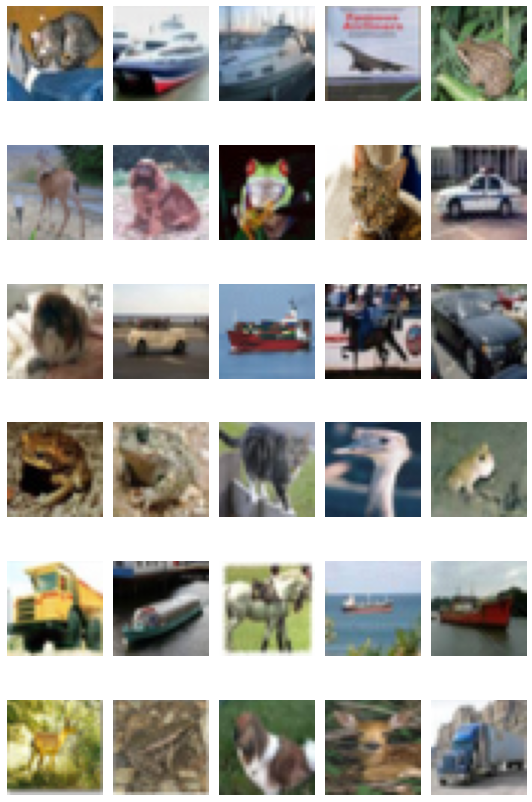}&
\includegraphics[height =3.5in, width = 2in]{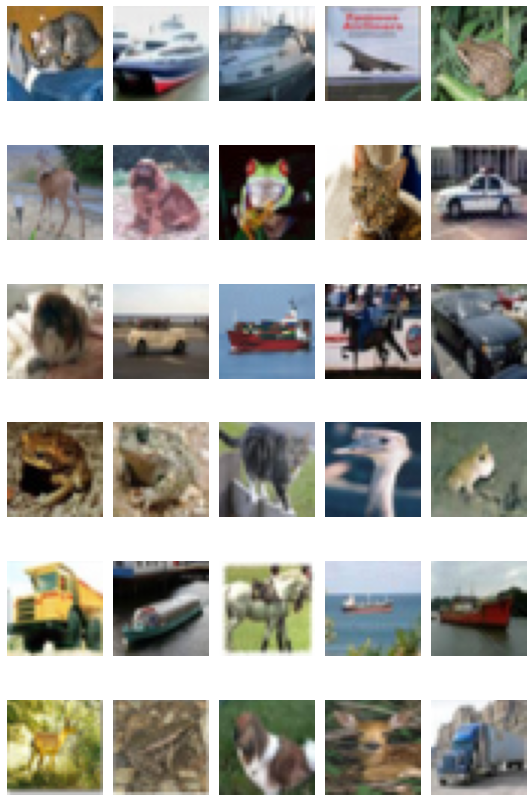}\\
PGD\_$l2$ & {\bf{PGAP}} & {\bf{FPGAP}} \\

\end{tabular}
\caption{Adversarial examples generated using multiple methods with similar fooling rate on CIFAR-10 dataset.}

\label{fig:cifar_all} 
\end{figure*}


\begin{figure*}
\centering
\subfloat[]{\includegraphics[width = 3.25in]{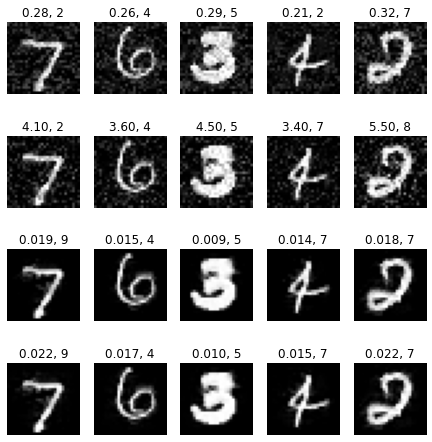}}
\hfil
\subfloat[]{\includegraphics[width=3.25in]{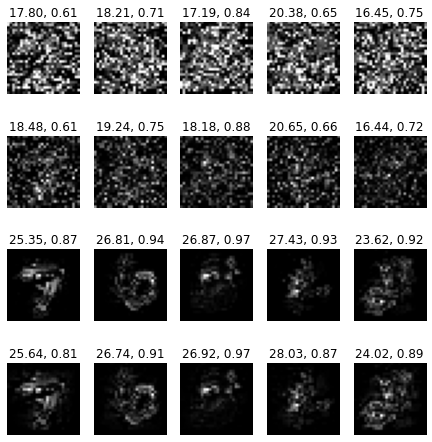}}
\caption{Image on the left: Adversarial examples generated by multiple methods: from top to bottom PGD, PGD$\_l2$, PGAP and FPGAP at fixed number of iterations (10). Numbers on top of each image are $\epsilon$ and prediction of adversarial image respectively. Image on the right: Absolute difference maps with respect to original image.  Numbers on top of each map are PSNR and SSIM.}
\label{fig:mnist_psnr_10}
\end{figure*}

\begin{figure*}
\centering
\subfloat[]{\includegraphics[width=3.25in]{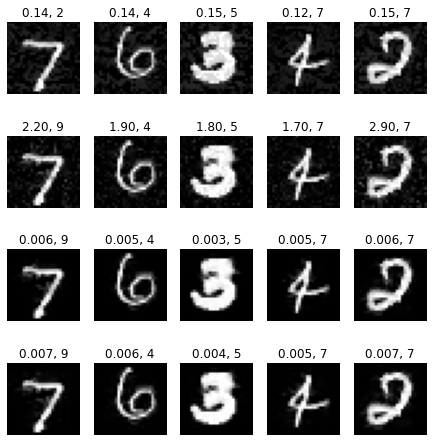}}
  \subfloat[]{\includegraphics[width=3.25in]{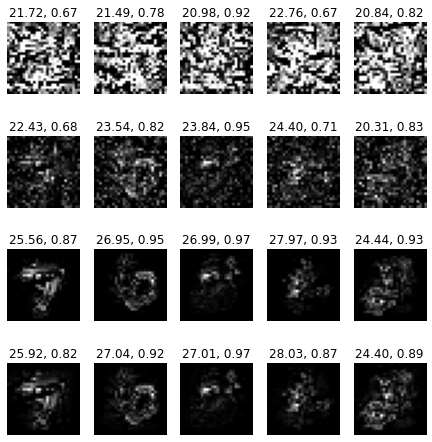}}
\caption{Image on the left: Adversarial examples generated by multiple methods: from top to bottom PGD, PGD$\_l2$, PGAP and FPGAP at fixed number of iterations (30). Numbers on top of each image are $\epsilon$ and prediction of adversarial image respectively. Image on the right: Absolute difference maps with respect to original image.  Numbers on top of each map are PSNR and SSIM.}
\label{fig:mnist_psnr_30}
\end{figure*}

\begin{figure*}
\centering
  \subfloat[]{\includegraphics[width=3.25in]{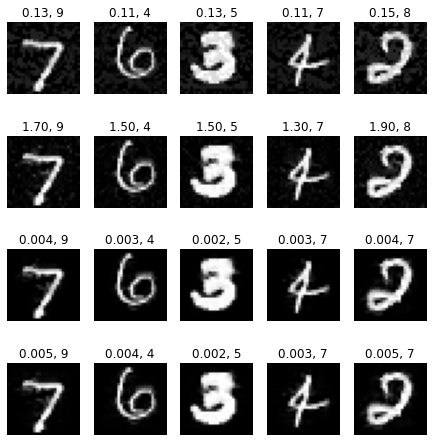}}
  \subfloat[]{\includegraphics[width=3.25in]{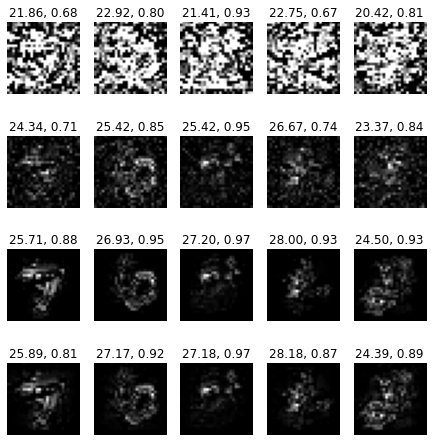}}
\caption{Image on the left: Adversarial examples generated by multiple methods: from top to bottom PGD, PGD$\_l2$, PGAP and FPGAP at fixed number of iterations (50). Numbers on top of each image are $\epsilon$ and prediction of adversarial image respectively. Image on the right: Absolute difference maps with respect to original image.  Numbers on top of each map are PSNR and SSIM.}
\label{fig:mnist_psnr_50}
\end{figure*}

\begin{figure*}
\centering
\subfloat[]{\includegraphics[width=3.25in]{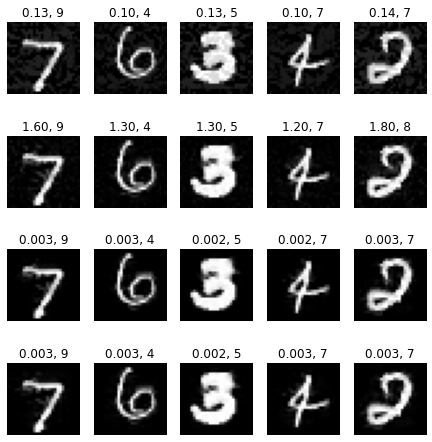}}
  \subfloat[]{\includegraphics[width=3.25in]{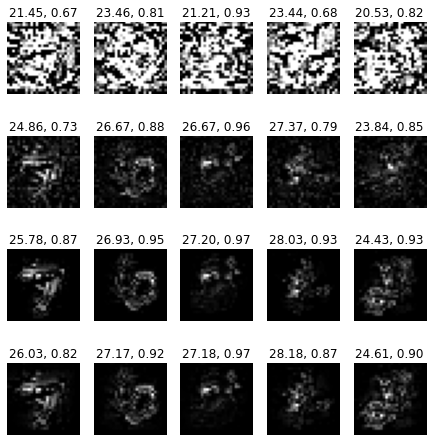}}
\caption{Image on the left: Adversarial examples generated by multiple methods: from top to bottom PGD, PGD$\_l2$, PGAP and FPGAP at fixed number of iterations (70). Numbers on top of each image are $\epsilon$ and prediction of adversarial image respectively. Image on the right: Absolute difference maps with respect to original image.  Numbers on top of each map are PSNR and SSIM.}
\label{fig:mnist_psnr_70}
\end{figure*}


\begin{figure*}
   \centering
\begin{tabular}{cc}

\includegraphics[width = 3.2in]{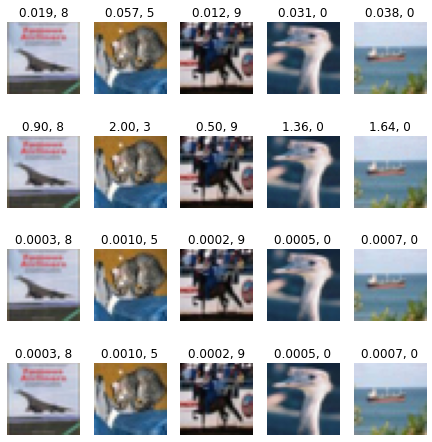}& 
\includegraphics[width = 3.2in]{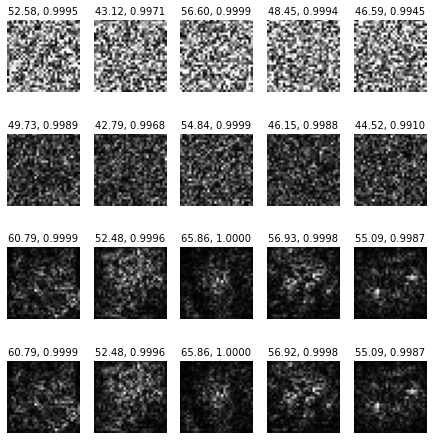}\\
(a) Adversarial examples &(b) Absolute difference map of R channel\\

\includegraphics[width = 3.2in]{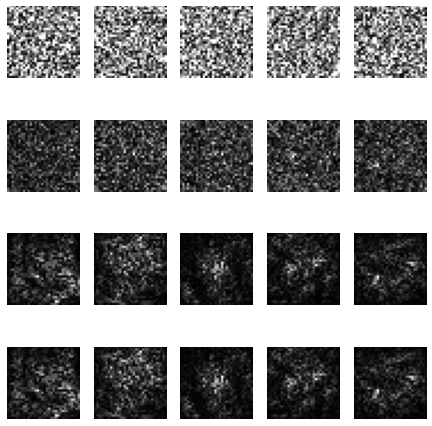}&
\includegraphics[width = 3.2in]{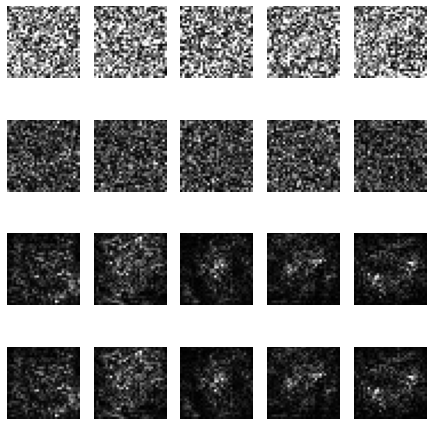}\\
(c) Absolute difference map of G channel &(d) Absolute difference map of B channel\\

\end{tabular}
\caption{(a): Adversarial examples generated by multiple methods: from top to bottom PGD, PGD$\_l2$, PGAP and FPGAP at fixed number of iterations (10). Numbers on top of each image are $\epsilon$ and prediction of adversarial image respectively. (b): Absolute difference maps of channel R with respect to original image, Numbers on top of each map are PSNR and SSIM. (c): Absolute difference maps of channel G with respect to original image. (d): Absolute difference maps of channel B with respect to original image}

\label{fig:cifar_psnr_10} 
\end{figure*}

\begin{figure*}
   \centering
\begin{tabular}{cc}

\includegraphics[width = 3.2in]{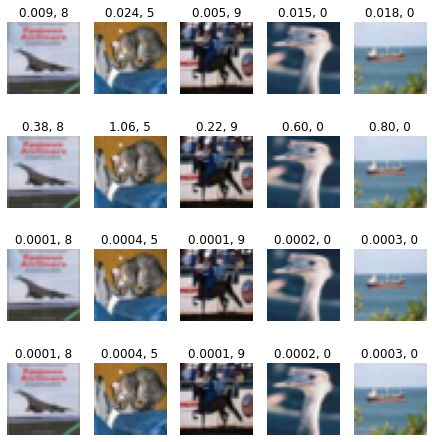}& 
\includegraphics[width = 3.2in]{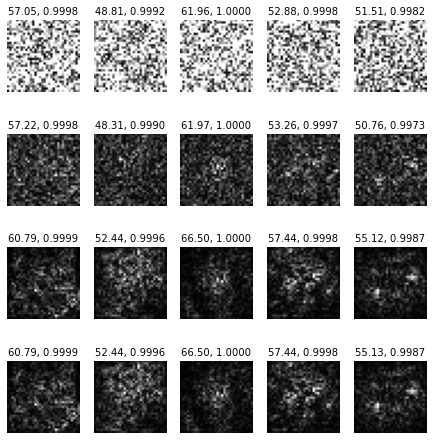}\\
(a) Adversarial examples &(b) Absolute difference map of R channel\\

\includegraphics[width = 3.2in]{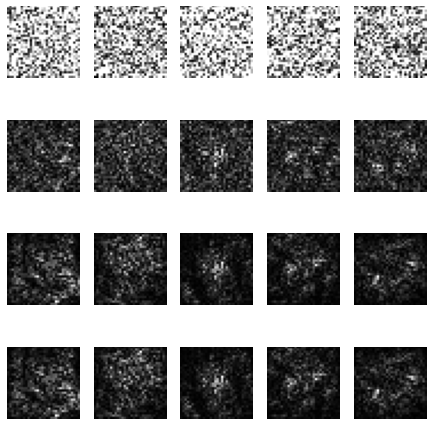}&
\includegraphics[width = 3.2in]{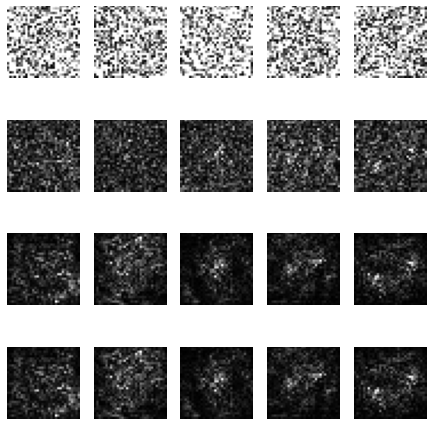}\\
(c) Absolute difference map of G channel &(d) Absolute difference map of B channel\\

\end{tabular}
\caption{(a): Adversarial examples generated by multiple methods: from top to bottom PGD, PGD$\_l2$, PGAP and FPGAP at fixed number of iterations (30). Numbers on top of each image are $\epsilon$ and prediction of adversarial image respectively. (b): Absolute difference maps of channel R with respect to original image, Numbers on top of each map are PSNR and SSIM. (c): Absolute difference maps of channel G with respect to original image. (d): Absolute difference maps of channel B with respect to original image}

\label{fig:cifar_psnr_30} 
\end{figure*}

\begin{figure*}
   \centering
\begin{tabular}{cc}

\includegraphics[width = 3.2in]{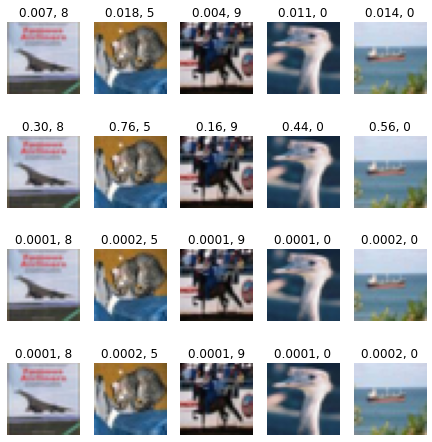}& 
\includegraphics[width = 3.2in]{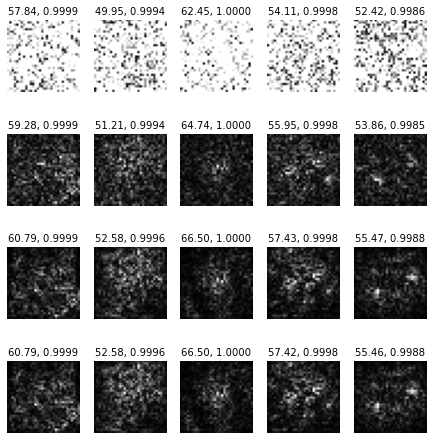}\\
(a) Adversarial examples &(b) Absolute difference map of R channel\\

\includegraphics[width = 3.2in]{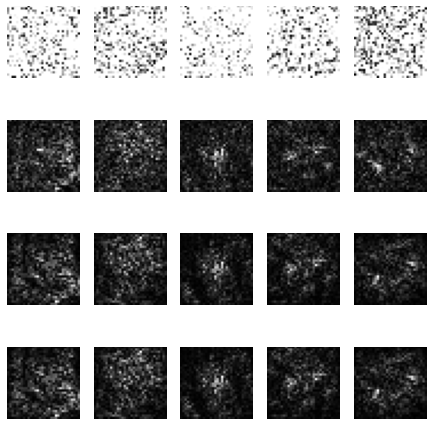}&
\includegraphics[width = 3.2in]{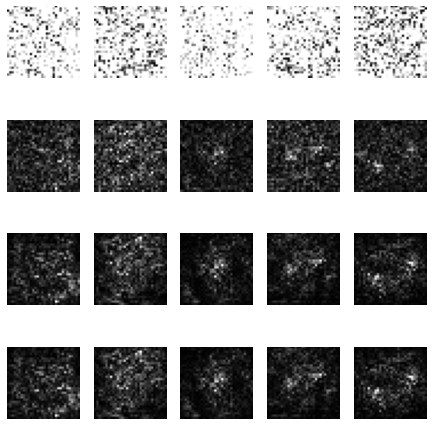}\\
(c) Absolute difference map of G channel &(d) Absolute difference map of B channel\\

\end{tabular}
\caption{(a): Adversarial examples generated by multiple methods: from top to bottom PGD, PGD$\_l2$, PGAP and FPGAP at fixed number of iterations (50). Numbers on top of each image are $\epsilon$ and prediction of adversarial image respectively. (b): Absolute difference maps of channel R with respect to original image, Numbers on top of each map are PSNR and SSIM. (c): Absolute difference maps of channel G with respect to original image. (d): Absolute difference maps of channel B with respect to original image}
\label{fig:cifar_psnr_50}
\end{figure*}

\begin{figure*}
   \centering
\begin{tabular}{cc}

\includegraphics[width = 3.2in]{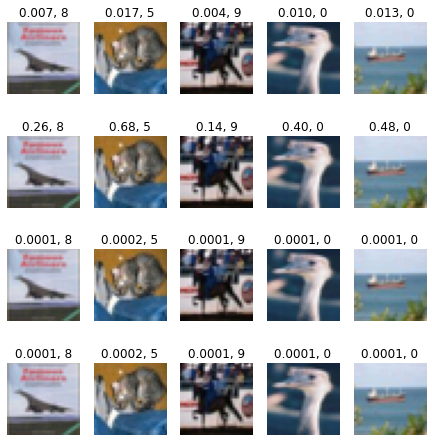}& 
\includegraphics[width = 3.2in]{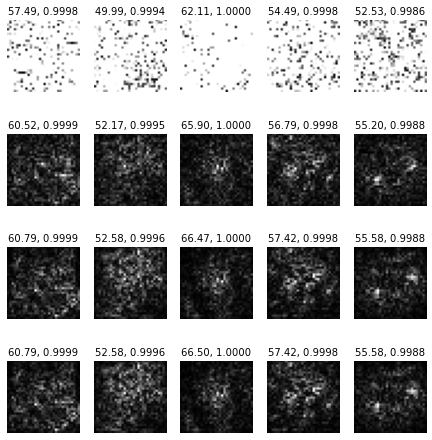}\\
(a) Adversarial examples &(b) Absolute difference map of R channel\\

\includegraphics[width = 3.2in]{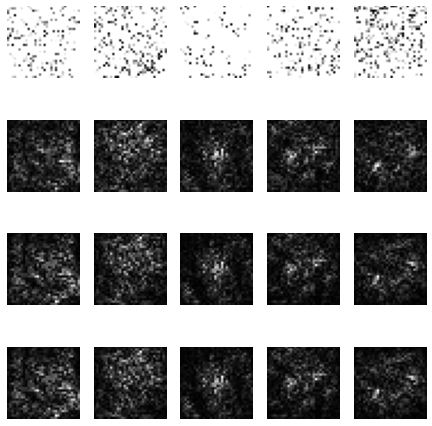}&
\includegraphics[width = 3.2in]{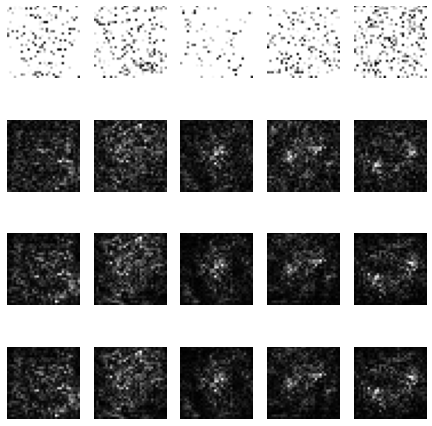}\\
(c) Absolute difference map of G channel &(d) Absolute difference map of B channel\\

\end{tabular}
\caption{(a): Adversarial examples generated by multiple methods: from top to bottom PGD, PGD$\_l2$, PGAP and FPGAP at fixed number of iterations (70). Numbers on top of each image are $\epsilon$ and prediction of adversarial image respectively. (b): Absolute difference maps of channel R with respect to original image, Numbers on top of each map are PSNR and SSIM. (c): Absolute difference maps of channel G with respect to original image. (d): Absolute difference maps of channel B with respect to original image}

\label{fig:cifar_psnr_70} 
\end{figure*}

\end{document}